# Study of polarization-dependent band inversions and edge states from two-dimensional square lattice plasmonic crystals

T.H. Chan*, Y.H. Guan*, C. Liu, and H.C. Ong†

Department of Physics, The Chinese University of Hong Kong, Shatin, Hong Kong, People's Republic of China

When two topologically trivial and nontrivial systems are brought together, a localized energy state is formed at the interface. For crystalline quantum and classical systems, their topology can be determined by studying the eigenmode symmetries at high symmetry points (HSPs). As electromagnetic systems are usually leaky, their radiations retain the eigenmode information, thus providing a means for probing the band topology. Here, we formulate the far-field characteristics of the eigenmodes at HSPs in 2D square lattice plasmonic systems and reveals, unlike the conventional tight-binding model, several polarization-dependent band inversions occur at the Γ and X points, rendering subtle changes in their band topology. In particular, by carefully tuning the system geometry to facilitate trivial and non-trivial phases, polarization selective 1D propagating and non-propagating edge states that possess distinct field symmetries are realized. The propagation of 1D edge state depends largely on the degree of bulk-edge interference. We perform angle- and polarization-resolved diffraction spectroscopy on 2D Au plasmonic nanohole arrays to verify the theory and observe such states. Our study demonstrates the applicability of simple far-field technique for diagnosing the band topology of various crystalline systems in electromagnetics and acoustics.

*These authors contributed equally to this work

† hcong@phy.cuhk.edu.hk

# I. INTRODUCTION

Topology has gone far beyond a mathematical concept and become a vital tool in studying condensed matter physics [1-4], mechanics [5,6], acoustics [7-9], electromagnetism [10-14] and even electronics [15,16]. All of them share a common feature that topologically protected edge state can be formed at the interface between two topologically trivial and nontrivial systems [1-16]. Such state is gaining a widespread attention and becoming a good candidate for making electronic, acoustic, and photonic devices [17-20].

In general, the topological order of a photonic system can be classified whether it possesses time-reversal symmetry or not, which thereby is characterized by zero or nonzero Chern number [10-14]. The most notable examples are gyromagnetic photonic crystals in which the applied magnetic field breaks the time-reversal symmetry emulating integer quantum Hall effect [21]. The magnetization results in an effective gauge field, leading to finite Berry curvature that makes the optical fields propagate in real space as if they are influenced by a momentum-space equivalent magnetic field [22]. As a result, nonzero Chern number yields topologically protected edge state following bulk-edge correspondence [21]. On the other hand, for those which respect time-reversal symmetry such as crystalline systems, despite their vanishing Chern number, various spatial symmetries further enrich classification, leading to optical analogous quantum spin Hall and valley Hall effects that have been captured intensive attention lately in optical frequencies where magnetization is difficult [13-15,23-25]. The spatial symmetries displace or obstruct the Wannier functions, facilitating three kinds of topological bands, which are obstructed atomic limit (OAL), fragile, and topologically stable, depending on the "Wannierizability" of the Wannier functions [26]. Both OAL and fragile bands do not produce non-Wannierzible functions and thus are not stable [27].

Many two-dimensional (2D) topological photonic systems studied so far are based on crystalline symmetry. For example, for 2D Su-Schrieffer-Heeger (SSH) square lattices that respect both time-reversal and inversion symmetries, although their Berry curvature is zero, they still exhibit nontrivial topological phase by considering 2D vectorial Zak phase [28,29]. A system with $(\pi,\pi)$ Zak phase, which is featured by the band inversions at the X and Y points while those at the $\Gamma$ and M points remain unchanged, is defined as topologically nontrivial and 1D edge state and even higher order 0D corner state can be realized accordingly [30,31]. In fact, 2D SSH model is a typical OAL phase in which the Wannier functions are shifted away from the center maximal Wyckoff position for trivial to the corner for nontrivial [26]. Other than SSH model, different hexagonal lattices have been realized that exploit pseudo-spin and valley degrees of freedom, manifesting OAL and fragile phases [26,27].

For crystalline photonic systems, it is essential to characterize their band topologies in prior to designing the associated edge state. Unfortunately, measuring band topology is a challenging task as it involves sophisticated instrumentations that are difficult to implement [32-39]. Recently, the development of topological quantum chemistry and symmetry indicators may simplify the task by exploiting the eigenmode symmetries at high symmetry points (HSPs) [26,27,40-43]. The symmetry-indicator invariants below the band gap of interest can then be defined, subsequently yielding the positions of the Wannier functions [26]. Since most of the photonic systems are leaky, the eigenmode symmetries are hidden in the continuum that could be extracted. However, such extraction is not trivial considering so many physical observables radiations can have. In fact, a few studies have been attempted but they are all limited to the lowest Γ point based on simple symmetry consideration and not yet extended to higher HSPs [44-48]. Therefore, it is desired to develop a systematic approach to measure the optical properties of the eigenmodes at HSPs.

In this work, we employ temporal coupled mode theory (CMT) to study the irreducible representations (irrep) of 2D square lattice Au nanohole plasmonic crystals (PmCs). It shows the band topologies of the PmCs with different circular hole sizes can be determined directly by far-field. In particular, by combining finite-difference time-domain (FDTD) method and angle- and polarization-resolved diffraction spectroscopy, we find the band inversions at Γ and X points are not only hole size but also polarization dependent, giving rise to trivial and nontrivial 2D Zak phases. By carefully choosing two PmCs with distinctive Zak phases to form heterostructure, we observe polarization selective 1D propagating and non-propagating edge states, as evident from simulation and experiment. While s-excited edge state propagates along the interface, p-excited edge state is stationary. Their propagation is largely governed by the overlapping between the edge and bulk states. Although square lattice with inversion symmetry is demonstrated here, we anticipate our approach can be further generalized to different lattices and bases carrying different symmetries [26].

## II. METHOD

We first use CMT to illustrate how the irrep of the eigenmodes at HSPs can be determined by far-fields [49-52]. We consider an optically thick 2D square lattice PmC that supports Bloch-like surface plasmon polariton (SPP) modes and its dispersion relation is given by the phase-matching equation as $\vec{k}_{SPP} = \left(\frac{2\pi}{\lambda}\sin\theta\cos\varphi + \frac{n_x 2\pi}{P}\right)\hat{x} + \left(\frac{2\pi}{\lambda}\sin\theta\sin\varphi + \frac{n_y 2\pi}{P}\right)\hat{y}$ , where θ is the incident polar angle, φ is the azimuthal angle defined with respect to the Γ-X direction, λ is the wavelength, P is the period of the lattice, $n_x$, and $n_y$ are the indices specifying the Bragg scattering order, and $\left|\vec{k}_{SPP}\right| = \frac{2\pi}{\lambda}\sqrt{\frac{\varepsilon_{Au}}{1+\varepsilon_{Au}}}$ is the propagation constant with $\varepsilon_{Au}$ being the Au dielectric constant [42]. As SPPs are nonradiative evanescent waves with in-plane momentum larger than that of light, the phase-matching equation folds the SPP dispersion relation into the light cone, making the modes above it excitable by far-field [53]. Several dispersive ($n_x$,$n_y$) SPP modes for P = 750 nm are plotted in Fig. 1(a) along Γ-X (φ = 0°), Γ-M (φ = 45°), and X-M (θ = $\sin^{-1}$(λ/2P)°) directions, and they are labeled as (±1,0), (0,±1), and (-1,±1) SPPs. We expect energy band gaps will emerge between the modes at HSPs due to mode coupling. For example, the lowest energy gap at the Γ point, indicated by the red dash box, is formed by the coupling between (±1,0) and (0,±1) SPPs. The second lowest gap at the X point, indicated by the blue dash box, is due to the coupling between (0,±1) and (-1,±1) SPPs. Finally, as given in the Supplementary Information [54], the lowest gap at the M point is below the light cone and it arises from the coupling between (0,0), (-1,0), (0,-1), and (-1,-1) SPPs. Therefore, the SPP modes at the Γ and X gaps are accessible by far-fields but not those at the M gap, which are near-field excited. It will be shown in the following that the band topologies are dictated by the eigenmodes at Γ and X points [26,28-31], they are thus being focused.

We then perform FDTD simulations to verify the CMT results as well as to illustrate polarization-dependent band inversions. A series of 2D semi-infinitely thick Au square lattice nanohole arrays with P = 750 nm, hole height H = 100 nm, and hole radius R varying from 150 to 300 nm are excited by polarized plane wave and dipole excitations to simulate their far-field diffraction spectra and near-field patterns [54].

Finally, we measure 2D Au PmCs by polarization- and θ-resolved diffraction spectroscopy along Γ-X and Γ-M directions [54,55]. 200 nm thick Au films are deposited by magnetron sputtering on c-axis sapphire substrate at room temperature and then patterned by focused ion beam (FIB). After sample preparation, the PmCs are transferred to a homebuilt Fourier space optical microscope described in the Supplementary Information [54].

## III. AT Γ POINT

### A. CMT

At the Γ point, the four propagating (±1,0) and (0,±1) SPPs are illustrated in Fig. 1(b), showing they propagate in the ±x and ±y directions deduced by the phase-matching equation with $\left|\vec{k}_{SPP}\right| = 2\pi/P$. The CMT equation, which has the general form of $\frac{dA}{dt} = iHA + KS_{+}$, can be expressed as [54,55]:

$$\frac{d}{dt}\begin{bmatrix} a_1 \\ a_2 \\ a_3 \\ a_4 \end{bmatrix} = i\begin{bmatrix} \tilde{\omega}_o & \alpha & \beta & \beta \\ \alpha & \tilde{\omega}_o & \beta & \beta \\ \beta & \beta & \tilde{\omega}_o & \alpha \\ \beta & \beta & \alpha & \tilde{\omega}_o \end{bmatrix}\begin{bmatrix} a_1 \\ a_2 \\ a_3 \\ a_4 \end{bmatrix} + \begin{bmatrix} \kappa & 0 \\ -\kappa & 0 \\ 0 & \kappa \\ 0 & -\kappa \end{bmatrix}\begin{bmatrix} s_{p+} \\ s_{s+} \end{bmatrix}, \quad (1)$$

where $a_{1-4}$ are the mode amplitudes, $\tilde{\omega}_o = \omega_o + i\Gamma_o/2$ is the complex angular frequency in which $\omega_o$ and $\Gamma_o$ are the resonant frequency and decay rate of SPPs. Because the Γ point possesses $C_{4v}$ symmetry [56], for the Hamiltonian H matrix, the complex coupling constants are defined as α and β for the modes propagating in the same and perpendicular directions, respectively. They are complex provided both near- and far-field interactions are present [52]. In addition, for the in-coupling matrix K and the incident power $S_+$, each mode supports only one input channel with an in-coupling constant defined as κ, and the incident power amplitudes are defined as $s_{p+}$ and $s_{s+}$ for p- and s-polarizations in positive x- and y-directions. The sign of κ depends on the relative orientation between the incident polarization and the transverse polarization of the SPP, which lies in the plane defined by $\vec{k}_{SPP}$, resulting in 0 or π phase shift [55]. Four channels will interfere and yield a specular (total) reflection. Once Eq. (1) is formulated, we then solve the eigenvalues and eigenvectors by diagonalization such that $T\frac{dA}{dt} = \frac{d\tilde{A}}{dt} = iTHT^{-1}TA + TKS_{+} = i\tilde{H}\tilde{A} + \tilde{K}S_{+}$, where T is the transformation matrix [54]. Eq. (1) can then be rewritten as:

$$\frac{d}{dt}\begin{bmatrix} \tilde{a}_1 \\ \tilde{a}_2 \\ \tilde{a}_3 \\ \tilde{a}_4 \end{bmatrix} = i\begin{bmatrix} \tilde{\omega}_o - \alpha & 0 & 0 & 0 \\ 0 & \tilde{\omega}_o - \alpha & 0 & 0 \\ 0 & 0 & \tilde{\omega}_o + \alpha - 2\beta & 0 \\ 0 & 0 & 0 & \tilde{\omega}_o + \alpha + 2\beta \end{bmatrix}\begin{bmatrix} \tilde{a}_1 \\ \tilde{a}_2 \\ \tilde{a}_3 \\ \tilde{a}_4 \end{bmatrix} - \sqrt{2}\begin{bmatrix} 0 & \kappa \\ \kappa & 0 \\ 0 & 0 \\ 0 & 0 \end{bmatrix}\begin{bmatrix} s_{p+} \\ s_{s+} \end{bmatrix}. \quad (2)$$

From Eq. (2), one sees from $\tilde{H}$ that $\tilde{a}_{1,2}$ modes are degenerate but $\tilde{a}_{3,4}$ modes are nondegenerate. Two band gaps thus are formed. The electric field patterns of $\tilde{a}_{1-4}$ eigenvectors are calculated in Fig. 1(c), showing the irreps of $\tilde{a}_{1,2}$ are E whereas $\tilde{a}_3$ and $\tilde{a}_4$ are $B_1$ and $A_1$ following the $C_{4v}$ character table [56]. We see from $\tilde{K}$ in Eq. (2) that E modes are radiative bright modes excited individually by s- and p-polarized lights. On the other hand, both $A_1$ and $B_1$ modes are symmetry protected bound states in the continuum (BICs) with zero in-coupling constants [57,58]. Although their angular frequencies are still complex due to the absorption loss, the modes have zero radiative decay rates and do not interact with far-field. To resolve $A_1$ and $B_1$, we notice from Fig. 1(c) that they exhibit different field symmetries after rotating the patterns by 45°. In other words, $B_1$ is asymmetric with respect to the Γ-M plane whereas $A_1$ is symmetric. Therefore, in the vicinity to the Γ point along the Γ-M direction where $C_{4v}$ symmetry is broken so that $A_1$ and $B_1$ no longer are BICs, $B_1$ is s-excited but $A_1$ is p-excited.

**B. FDTD**

Fig. 2(a) & (c) show the p-polarized total reflectivity spectral mappings taken along the Γ-X direction at $\theta = 0°$ (normal incidence) and 1° for different R. While $\theta = 0°$ is at the Γ point where the BICs are present, $\theta = 1°$ will break the BICs, forcing the modes to show up. One sees in Fig. 2(a) that only one reflection dip is observed, and its spectral position varies slowly with R. Several spectra for R = 150, 200, 250, and 300 nm are extracted in Fig. S4-7 in the Supplementary Information for reference [54]. On the other hand, for $\theta = 1°$ in Fig. 2(c), two additional dips are seen, suggesting they are BICs. By comparing two mappings, we conclude the mode at $\theta = 0°$ can be assigned as E whereas other two modes at $\theta = 1°$ are $A_1$ and $B_1$. More importantly, we notice $A_1$ and $B_1$ initially are higher than E but gradually shift down as R increases. They cross with E one by one at R = 220 and 270 nm, leading to two band inversions. Fig. 2(b) & (d) show the corresponding s-polarized reflectivity mappings for $\theta = 0°$ and 1°, in which only one reflection dip is observed. Several spectra are plotted in Fig. S4-7 [54]. Clearly, the mode is found to be the same as the E mode in p-polarization and thus is the second E mode. We then calculate the field patterns of the E modes in Fig. S4-7, and they agree with Fig. 1(c) [54].

Once the radiative E modes are resolved, we then proceed to assign $A_1$ and $B_1$. Fig. 2(e) & (f) show the p- and s-polarized reflectivity mappings taken at $\theta = 1°$ along the Γ-M direction. In agreement with CMT, two modes are excited by p- and s-polarized lights independently. The p- and s-excited modes are assigned as $A_1$ and $B_1$, respectively. Their field patterns in Fig. S4-7 agree with those in Fig. 1(c) [54]. Therefore, by carrying out polarization-dependent reflectivity spectroscopy along the Γ-X and Γ-M directions, one can resolve $A_1$, $B_1$, and E unambiguously.

**C. Experiment**

Three PmCs are measured and their scanning electron microscopy (SEM) images are shown in the insets of Fig. 3(b), (d) & (f), showing they have P = 753 nm, H = 102 nm, and R

= 150, 250, and 300 nm. They are chosen as they locate before and after two band inversions in Fig. 2. The p- and s-polarized specular reflectivity mappings of three PmCs taken along the Γ-X direction are illustrated in Fig. 3. The SPP dispersions are sketched as dash lines in Fig. 3(a) & (b) for visualization. While the (±1,0) SPPs are p-excited, two degenerate (0,±1) modes are p- and s-excited. They are labelled as symmetric $(0,\pm1)_s$ and asymmetric $(0,\pm1)_a$ SPPs, respectively, due to their distinctive field symmetries [59].

When approaching to the Γ point, the mode dispersions are found to be strongly R dependent. For R = 150 nm in Fig. 3(a) & (b), as indicated by the red dash line, we see the p-excited (-1,0) and s-excited $(0,\pm1)_a$ SPPs located at 784 nm are degenerate, and they are assigned as E. On the other hand, the (1,0) and $(0,\pm1)_s$ SPPs vanish at the Γ point, forming two BICs at 761 and 778 nm. It is noticed that the band gap between the (-1,0) and $(0,\pm1)_s$ SPPs is very small, making the $(0,\pm1)_s$ BIC difficult to be resolved. We will show later that a small gap size is detrimental to the formation of edge state. To assign the (1,0) and $(0,\pm1)_s$ SPPs properly, the Γ-M p- and s-polarized mappings are shown in Fig. 4(a) & (b). One sees the (1,0) and $(0,\pm1)_s$ SPPs are BICs at 0° but become visible when θ increases. Essentially, the (1,0) and $(0,\pm1)_s$ SPPs are p- and s-excited and thus are assigned as $A_1$ and $B_1$. Therefore, the irreps from high to low energy are $A_1$, $B_1$, and E, which agree with FDTD simulation.

For R = 250 nm, the Γ-X p- and s-mappings are shown in Fig. 3(c) & (d), clearly indicating two BICs, arising from the (±1,0) SPPs, are observed at 760 and 797 nm. In addition, the $(0,\pm1)_s$ and $(0,\pm1)_a$ SPPs at 774 nm are degenerate, revealing they are E modes by the red dash line. To assign the (±1,0) SPPs, we examine the Γ-M p- and s-mappings in Fig. 4(c) & (d). As the (1,0) and (-1,0) SPPs are p- and s-excited, they are assigned as $A_1$ and $B_1$. The descending order thus is $A_1$, E, and $B_1$, showing one band inversion occurs between E and $B_1$.

For R = 300 nm in Fig. 3(e) & (f), we notice the p-excited (1,0) and s-excited $(0,\pm1)_a$ SPPs are degenerate E modes at 767 nm, and the (-1,0) and $(0,\pm1)_s$ SPPs are two BICs at 774 and 791 nm. Fig. 4(e) & (f) show while the (-1,0) mode is s-excited, the $(0,\pm1)_s$ mode is p-excited, making them as $B_1$ and $A_1$. As a result, the irreps are E, $A_1$, and $B_1$ in which E and $A_1$ undergo the second band inversion.

## IV. AT X POINT

### A. CMT

For the X point taken along the Γ-X direction, the four (0,±1) and (-1,±1) SPPs are illustrated in Fig. 1(d) with $\vec{k}_{SPP} = \pm(\hat{x} \pm 2\hat{y})(\pi/P)$. We define the complex coupling constants to be η, γ, and σ for the modes oriented relatively to each other by 60°, 120°, and 180°, respectively. As the X point follows $C_{2v}$ symmetry in which only two right- and left-hand diffractions are supported, the CMT equation can be expressed as:

$$\frac{d}{dt}\begin{bmatrix} a_1 \\ a_2 \\ a_3 \\ a_4 \end{bmatrix} = i\begin{bmatrix} \tilde{\omega}_o & \sigma & \gamma & \eta \\ \sigma & \tilde{\omega}_o & \eta & \gamma \\ \gamma & \eta & \tilde{\omega}_o & \sigma \\ \eta & \gamma & \sigma & \tilde{\omega}_o \end{bmatrix}\begin{bmatrix} a_1 \\ a_2 \\ a_3 \\ a_4 \end{bmatrix} + \begin{bmatrix} -\kappa_{pL} & \kappa_{sL} & -\kappa_{pR} & \kappa_{sR} \\ \kappa_{pR} & -\kappa_{sR} & \kappa_{pL} & -\kappa_{sL} \\ -\kappa_{pL} & -\kappa_{sL} & -\kappa_{pR} & -\kappa_{sR} \\ \kappa_{pR} & \kappa_{sR} & \kappa_{pL} & \kappa_{sL} \end{bmatrix}\begin{bmatrix} s_{pL+} \\ s_{sL+} \\ s_{pR+} \\ s_{sR+} \end{bmatrix}, \tag{3}$$

where the R and L subscripts in κ and $s_+$ represent the right and left incidences. After diagonalization, Eq. (3) is transformed as [54]:

$$\frac{d}{dt}\begin{bmatrix} \tilde{a}_1 \\ \tilde{a}_2 \\ \tilde{a}_3 \\ \tilde{a}_4 \end{bmatrix} = i\begin{bmatrix} \tilde{\omega}_o+\sigma-\gamma-\eta & 0 & 0 & 0 \\ 0 & \tilde{\omega}_o-\sigma+\gamma-\eta & 0 & 0 \\ 0 & 0 & \tilde{\omega}_o-\sigma-\gamma+\eta & 0 \\ 0 & 0 & 0 & \tilde{\omega}_o+\sigma+\gamma+\eta \end{bmatrix}\begin{bmatrix} \tilde{a}_1 \\ \tilde{a}_2 \\ \tilde{a}_3 \\ \tilde{a}_4 \end{bmatrix} + \tilde{K}\begin{bmatrix} s_{pL+} \\ s_{sL+} \\ s_{pR+} \\ s_{sR+} \end{bmatrix}, \tag{4}$$

where $\tilde{K} = \begin{bmatrix} 0 & \kappa_{sR}-\kappa_{sL} & 0 & -(\kappa_{sR}-\kappa_{sL}) \\ \kappa_{pR}+\kappa_{pL} & 0 & \kappa_{pR}+\kappa_{pL} & 0 \\ 0 & \kappa_{sR}+\kappa_{sL} & 0 & \kappa_{sR}+\kappa_{sL} \\ \kappa_{pR}-\kappa_{pL} & 0 & -(\kappa_{pR}-\kappa_{pL}) & 0 \end{bmatrix}$. We see all the coupled modes are nondegenerate and three band gaps are formed. We also see all modes show different polarization dependences, with $\tilde{a}_{1,3}$ are s-excited whereas $\tilde{a}_{2,4}$ are p-excited. The $\tilde{a}_{1\text{-}4}$ eigenvector field patterns are calculated in Fig. 1(e). Using the character table for $C_{2v}$ point group, we assign $\tilde{a}_1$ and $\tilde{a}_3$ to be $A_2$ and $B_2$ and $\tilde{a}_2$ and $\tilde{a}_4$ to be $B_1$ and $A_1$ [56].

To resolve them, we formulate their right and left diffractions. By using conservation of energy and time reversal symmetry, the outgoing radiation is expressed as $S_- = CS_+ + K^T T^T \tilde{A}$, where C is the non-resonant scattering matrix [49]. By taking $s_{sR+} \neq 0$ under s-polarization, the diffraction orders are:

$$\begin{bmatrix} s_{pL-} \\ s_{sL-} \\ s_{pR-} \\ s_{sR-} \end{bmatrix} = C\begin{bmatrix} 0 \\ 0 \\ 0 \\ s_{sR+} \end{bmatrix} + \begin{bmatrix} 0 \\ -(\kappa_{sR}-\kappa_{sL})^2 \\ 0 \\ (\kappa_{sR}-\kappa_{sL})^2 \end{bmatrix}\frac{s_{sR+}}{i(\omega-\tilde{\omega}_1)} + \begin{bmatrix} 0 \\ (\kappa_{sR}+\kappa_{sL})^2 \\ 0 \\ (\kappa_{sR}+\kappa_{sL})^2 \end{bmatrix}\frac{s_{sR+}}{i(\omega-\tilde{\omega}_3)}, \tag{5}$$

showing only two diffractions at $\tilde{\omega}_1 = \tilde{\omega}_o+\sigma-\gamma-\eta$ ($\tilde{a}_1$, $A_2$) and $\tilde{\omega}_3 = \tilde{\omega}_o-\sigma-\gamma+\eta$ ($\tilde{a}_3$, $B_2$) are present. In addition, while the left and right diffractions from $A_2$ are π out of phase, those from $B_2$ are in phase [55]. Likewise, for $s_{pR+} \neq 0$, we have

$$\begin{bmatrix} s_{pL-} \\ s_{sL-} \\ s_{pR-} \\ s_{sR-} \end{bmatrix} = C \begin{bmatrix} 0 \\ 0 \\ s_{pR+} \\ 0 \end{bmatrix} + \begin{bmatrix} \left(\kappa_{pR}+\kappa_{pL}\right)^2 \\ 0 \\ \left(\kappa_{pR}+\kappa_{pL}\right)^2 \\ 0 \end{bmatrix} \frac{s_{pR+}}{i\left(\omega-\tilde{\omega}_2\right)} + \begin{bmatrix} -\left(\kappa_{pR}-\kappa_{pL}\right)^2 \\ 0 \\ \left(\kappa_{pR}-\kappa_{pL}\right)^2 \\ 0 \end{bmatrix} \frac{s_{pR+}}{i\left(\omega-\tilde{\omega}_4\right)}, \tag{6}$$

indicating only p-polarized diffractions at $\tilde{\omega}_2 = \tilde{\omega}_o - \sigma + \gamma - \eta$ ($\tilde{a}_2$, $B_1$) and $\tilde{\omega}_4 = \tilde{\omega}_o + \sigma + \gamma + \eta$ ($\tilde{a}_4$, $A_1$) exist. The left and right diffractions are in phase for $B_1$ but π out of phase for $A_1$. Therefore, by measuring the diffractions under p- and s-incidences, we can identify $A_{1,2}$ and $B_{1,2}$.

### B. FDTD

We show the Γ-X p- and s-polarized total reflectivity spectral mappings taken at the X point as a function of R in Fig. 5(a) & (b). They indicate four nondegenerate modes are supported with a pair of modes excited under one polarization. These two pairs are $A_1$ and $B_1$ and $A_2$ and $B_2$ for the p- and s-mappings. Remarkably, we notice $A_1$($A_2$) and $B_1$($B_2$) undergo p(s)-polarization band inversion when R increases. The corresponding field patterns of all four modes are calculated in Fig. S9-12, and they agree well with Fig. 1(e) [54].

To resolve them, we first simulate the p- and s-polarized specular reflectivity ($r_0$) and -1[st] diffraction order ($r_{-1}$) spectra for R = 150, 200, 250, and 300 nm in Fig. 6 and then fit them by Eq. (5) & (6). To simplify the fitting, we rewrite Eq. (6) as two superimposed Lorentzian profiles as [54,55]:

$$\begin{bmatrix} r_{p,0} \\ r_{p,-1} \end{bmatrix} = \begin{bmatrix} p_{0,b} \\ ip_{-1,b} \end{bmatrix} + \begin{bmatrix} \dfrac{\delta}{i\left(\omega-\tilde{\omega}_+\right)} - \dfrac{\gamma}{i\left(\omega-\tilde{\omega}_-\right)} \\ d\left(\dfrac{\delta}{i\left(\omega-\tilde{\omega}_+\right)} + \dfrac{\gamma}{i\left(\omega-\tilde{\omega}_-\right)}\right) \end{bmatrix} \tag{7}$$

where $p_{0,b}$ and $p_{-1,b}$ are the nonresonant backgrounds, $\tilde{\omega}_+$ and $\tilde{\omega}_-$ are the complex angular frequencies of the coupled modes with $\mathrm{Re}(\tilde{\omega}_+) > \mathrm{Re}(\tilde{\omega}_-)$, and δ and γ are the constants. The symmetry factor d is either 1 or -1, which determines whether $\tilde{\omega}_+$ mode is $A_1$ or $B_1$. For example, if d is found to be 1, two diffraction orders from $\tilde{\omega}_+$ are in phase and thus $\tilde{\omega}_+$ is $\tilde{\omega}_2$ or $B_1$ whereas $\tilde{\omega}_-$ is $\tilde{\omega}_4$ or $A_1$. On the other hand, if d = -1, then two diffraction orders from $\tilde{\omega}_+$ are π out of phase, and $\tilde{\omega}_+$ and $\tilde{\omega}_-$ are $\tilde{\omega}_4$ ($A_1$) and $\tilde{\omega}_2$ ($B_1$), respectively. Likewise, Eq. (5) is rewritten as [54]:

$$\begin{bmatrix} r_{s,0} \\ r_{s,-1} \end{bmatrix} = \begin{bmatrix} s_{0,b} \\ is_{-1,b} \end{bmatrix} + \begin{bmatrix} \dfrac{\mu}{i(\omega-\tilde{\omega}_+)} + \dfrac{\varepsilon}{i(\omega-\tilde{\omega}_-)} \\ d\left(\dfrac{\mu}{i(\omega-\tilde{\omega}_+)} - \dfrac{\varepsilon}{i(\omega-\tilde{\omega}_-)}\right) \end{bmatrix} \tag{8}$$

where $s_{0,b}$ and $s_{-1,b}$ are the nonresonant backgrounds and μ and ε are the constants. $\tilde{\omega}_+$ is $\tilde{\omega}_3$ ($B_2$) when d = 1 but is $\tilde{\omega}_1$ ($A_2$) for d = -1.

We then fit the spectra in Fig. 6(e)-(l) by using $\left|r_{p/s,0}\right|^2$ and $\left|r_{p/s,-1}\right|^2$ from Eq. (7) & (8) and the best fits are shown as dash lines. The best fitted parameters are tabulated in Table 1, showing d gradually flips from -1(-1) to 1(1) for p(s)-polarization when R increases. It indicates, for p-polarization, $A_1$ is higher than $B_1$ for small R but switch places as R increases. They undergo a band inversion at around R ~ 250 nm. Such flip is consistent with the field patterns in Fig. S9-12 [54]. Likewise, for s-polarization, $A_2$ is higher than $B_2$ for small R but go through a band inversion also at R ~ 230 nm. The irreps, $A_{1,2}$ and $B_{1,2}$, are labelled in Fig. 3(a) & (b).

## C. Experiment

We examine the Γ-X p- and s-polarized specular reflectivity mappings of the PmCs taken around the X point in Fig. 7(a)-(f). The dispersion relations are sketched in Fig. 7(a) & (b), showing two p-excited $(0,\pm1)_s$ and $(-1,\pm1)_s$ SPPs that are $A_1$ and $B_1$ and s-excited $(0,\pm1)_a$ and $(-1,\pm1)_a$ that are $A_2$ and $B_2$. All of them are nondegenerate. We see the crossing of $(0,\pm1)_s$ and $(-1,\pm1)_s$ as well as $(0,\pm1)_a$ and $(-1,\pm1)_a$ yield a band gap at ~ θ = 25°. Again, similar to those at the Γ point, the dispersions of three PmCs near the X point are strongly R dependent.

For R = 150 nm, from Fig. 7(a) & (b), they show the p(s)-polarized higher and lower SPPs are at 682(698) and 700(715) nm. We then plot their p- and s-polarized specular and -1$^{st}$ diffraction spectra in Fig. 8(a) & (b) and fit them by using Eq. (7) & (8) with the best fits displayed as dash lines. The fitted parameters, tabulated in Table 2, reveals d = -1 for both p- and s-polarizations. Therefore, the p(s)-excited higher and lower energy modes are assigned as $A_1$($A_2$) and $B_1$($B_2$).

On the other hand, for R = 250 nm, the p(s)-excited higher and lower energy modes are at 690(687) and 707(710) nm in Fig. 7(c) & (d). Their p- and s-polarized diffraction spectra and the best fits are shown in Fig. 8(c) & (d). Table 2 shows d = 1 for two polarizations, deducing the p(s)-polarized higher and lower energy modes are $B_1$($B_2$) and $A_1$($A_2$) after band inversion. Finally, for R = 300 nm, the p(s)-excited higher and lower energy modes are at 696 (677) and 715 (706) nm in Fig. 7(e) & (f). The corresponding specular reflectivity and -1$^{st}$ diffraction order spectra and the best fits are displayed in Fig. 8(e) & (f). Table 2 shows their d are 1 and thus the p(s)-excited higher and lower energy modes remain as $B_1(B_2)$ and $A_1(A_2)$.

## V. AT M POINT

As illustrated in Fig. 1(f), four (0,0), (-1,0), (0,-1) and (-1,-1) SPPs couple at the M point with $\vec{k}_{SPP} = \pm(\hat{x} \pm \hat{y})\pi/P$, similar to those of the Γ point after 45° rotation [54]. Therefore, the functional forms of $\tilde{H}$ and $\tilde{A}$ remain to be the same. The field patterns of $\tilde{a}_{1\text{-}4}$ are shown in Fig. 1(g). From Fig. S13 in the Supplementary Information, we observe no band inversion from R = 150 to 300 nm [54]. The order from high to low energy is $A_1$, $B_2$ and E.

## VI. REALIZATION OF POLARIZATION SELECTIVE EDGE STATES

Once the irreps are assigned, we are in the position of realizing polarization-dependent edge states. As an example, we examine R = 150 and 300 nm PmCs and study their 2D polarizations P. For $C_{4v}$ systems with inversion symmetry, P = $(P_x, P_y)$ with $P_x = P_y$, and $P_x$ is defined as $\frac{1}{2}\left(\sum_n q_n \bmod 2\right)$, where $(-1)^{q_n} = \frac{\eta_n(X)}{\eta_n(\Gamma)}$, $\sum_n q_n$ is the sum of $q_n$ of all n bands below the band gap of interest, and $\eta_n$ is the parity at the X or Γ point of the n[th] band [28,29,60,61]. $P_x$ is 0 or 1/2, yielding 0 and π Zak phase. It is noted that despite the systems are non-Hermitian, the Zak phase still can be quantized as 0 and π provided they possess inversion symmetry [62]. The dispersion relations for 150 and 300 nm PmCs are sketched in Fig. 9(a) & (b), where the energy bands are numbered and the corresponding eigenmodes, $A_{1,2}$, $B_{1,2}$, and E, are labelled at the Γ and X points. We then tabulate their parities, even (+) and odd (−), based on the irreps in Fig. 9(c). The blue dash lines indicate the band gaps which are opened by A and B modes but not degenerate E modes at the X and M points. For those modes that are not far-field excited, their parities are calculated by FDTD. The corresponding P for the 1[st], 2[nd], and 3[rd] gaps are then determined in Fig. 9(c), revealing the band topologies. Depending on the spectral alignments of the gaps between two PmCs, the heterostructure may support edge state. For example, the 1[st] and 2[nd] gaps above band 3 as well as the 3[rd] gaps above band 5 in 150 and 300 nm PmCs are aligned, and they exhibit trivial and nontrivial P, thus could yield edge states.

To demonstrate it, we simulate a supercell of the 150/300 PmC heterostructure as shown in Fig. 10(a), which consists of 20 unit cells on the right- and left-hand sides of the interface. The interface is along the y-direction. Fig. 10(b) & (c) show Γ-X p- and s-polarized reflectivity mappings of the heterojunction taken near the Γ and X points. We see the reflection dips at 778 and 707 nm, as indicated by red arrows, are localized within the band gaps, suggesting they are edge states [63]. The corresponding near-field patterns close to the interface are illustrated in Fig. 10(b) and (d), clearly showing the fields are strongly localized at the interface and decay rapidly. In addition, we observe the edge states possess different field symmetries, where the p-excited state is symmetric, but the s-excited state is asymmetric.

We then resolve whether the edge states are localized or propagating in the y-direction. Fig. 11(a) shows the $k_y$-dependent radiation spectral mapping excited by linear dipole being placed at the interface to effectively excite the edge states. $k_y$ is normalized with $P/2\pi$. At $k_y = 0$ in the Γ-X direction, we see two modes at 778 and 707 nm matching well with the p- and s-edge states in Fig. 10(b) & (c). For $k_y > 0$, the field patterns of the p-state taken at $(\lambda,k_y) = (783,0.02)$ and $(803,0.04)$ are illustrated in Fig. 11(b) & (c). Unlike Fig. 10(d), they are bulk states as the fields spread over the bulk regions, indicating the edge and bulk states interfere considerably primarily due to the small band gap as discussed earlier [64,65]. As the edge state must be isolated from the bulk band gap to maintain topological robustness, our results suggest the edge state does not propagate in the y-direction, making it stationary. For the s-edge state, the field patterns taken at $(\lambda,k_y) = (699,0.02)$ and $(710,0.02)$ for the upper and lower bands are shown in Fig. 11(d) & (e). Interestingly, while the upper band retains the localized field feature, implying it is a 1D propagating edge state, the lower band indicates the edge state is mixed with the bulk, losing its topological protection by increasing scattering to the bulk [64,65]. The upper band remains as an edge state even at (688,0.08) in Fig. 11(f) but starts experiencing scattering. More field patterns at different $(\lambda,k_y)$ of the upper and lower bands are given in the Supplementary Information for reference [44]. Nevertheless, the topological protection can be improved if one can widen up the gap size to reduce bulk-edge overlapping.

Finally, we fabricate the heterojunction by FIB and its SEM image is shown in Fig. 12(a). The circular hole arrays have P = 750 nm, H = 100 nm, and R = 150 and 300 nm. Their p- and s-polarized angular reflectivity mappings taken at the Γ and X points are shown in Fig. 12(b) & (c). Two localized modes are observed at 777 and 694 nm, demonstrating the presence of the edge states.

## VII. CONCLUSION

In summary, we have determined the irreps of the energy bands at the HSPs in the Brillouin zone of 2D square lattice plasmonic systems. By using temporal coupled mode theory (CMT), we devise a far-field approach that enables us to resolve the field symmetries of the photonic modes accordingly. Both our numerical and experimental results based on angle- and polarization-resolved diffraction spectroscopy agree very well with CMT. Our results demonstrate, for 2D nanohole plasmonic arrays with different hole radii, the modes at the Γ and X points undergo subtle polarization-dependent band inversions. The band inversions are then utilized for realizing p- and s-polarized 1D non-propagating and propagating edge states. Our study poses new ways in probing the topological properties of non-Hermitian electromagnetic systems for designing and realizing topologically protected edge and corner states.

## VIII. ACKNOWLEDGEMENT

This research was supported by the Chinese University of Hong Kong through Innovative Technology Fund Guangdong-Hong Kong Technology Cooperation Funding Scheme (GHP/077/20GD), Partnership Research Program (PRP/048/22FX), and Seed Program (ITS/245/23).

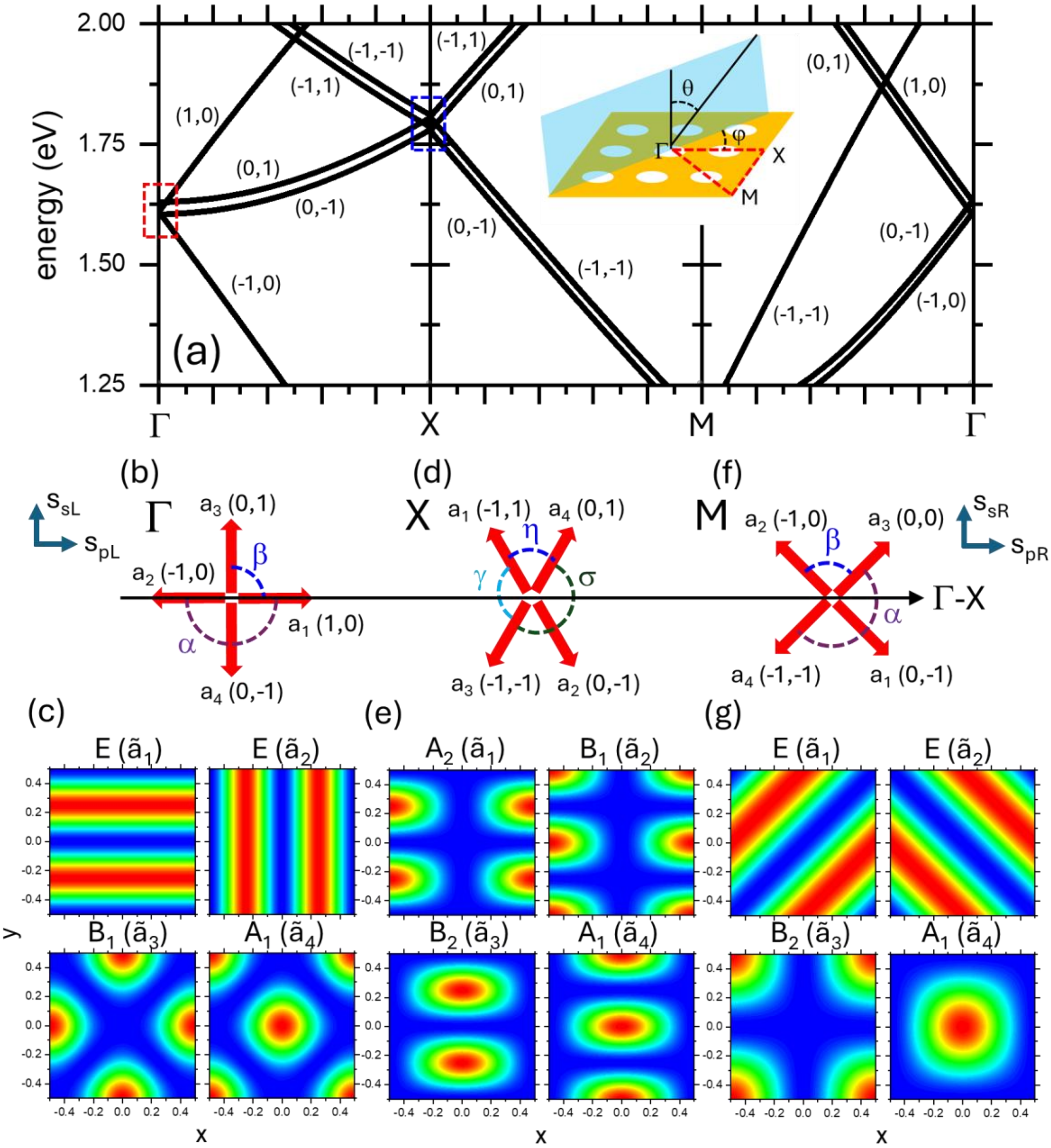

Fig. 1. (a) The dispersion relations of 2D square lattice PmC in the Brillouin zone. The double degeneracy is offset slightly for visualization. Energy band gaps are expected to form at the Γ and X points where SPPs couple. Inset: the square lattice and the excitation condition. (b) At the Γ point, the schematic of how four (±1,0) and (0,±1) SPPs couple and the coupling constants are defined as α and β. (c) The CMT deduced field patterns of $\tilde{a}_{1\text{-}4}$ and are labelled as E, $B_1$, and $A_1$. (d) At the X point, the schematic of how four (0,±1) and (-1,±1) SPPs couple and the coupling constants are defined as η, γ, and σ. (e) The field patterns of $\tilde{a}_{1\text{-}4}$ and are labelled as $A_2$, $B_1$, $B_2$, and $A_1$. (f) At the M point, the schematic of how four (0,0), (-1,0), (0,-1) and (-1,-1) SPPs couple and the coupling constants are defined as α and β. (g) The field patterns of $\tilde{a}_{1\text{-}4}$ and are labelled as E, $B_2$, and $A_1$.

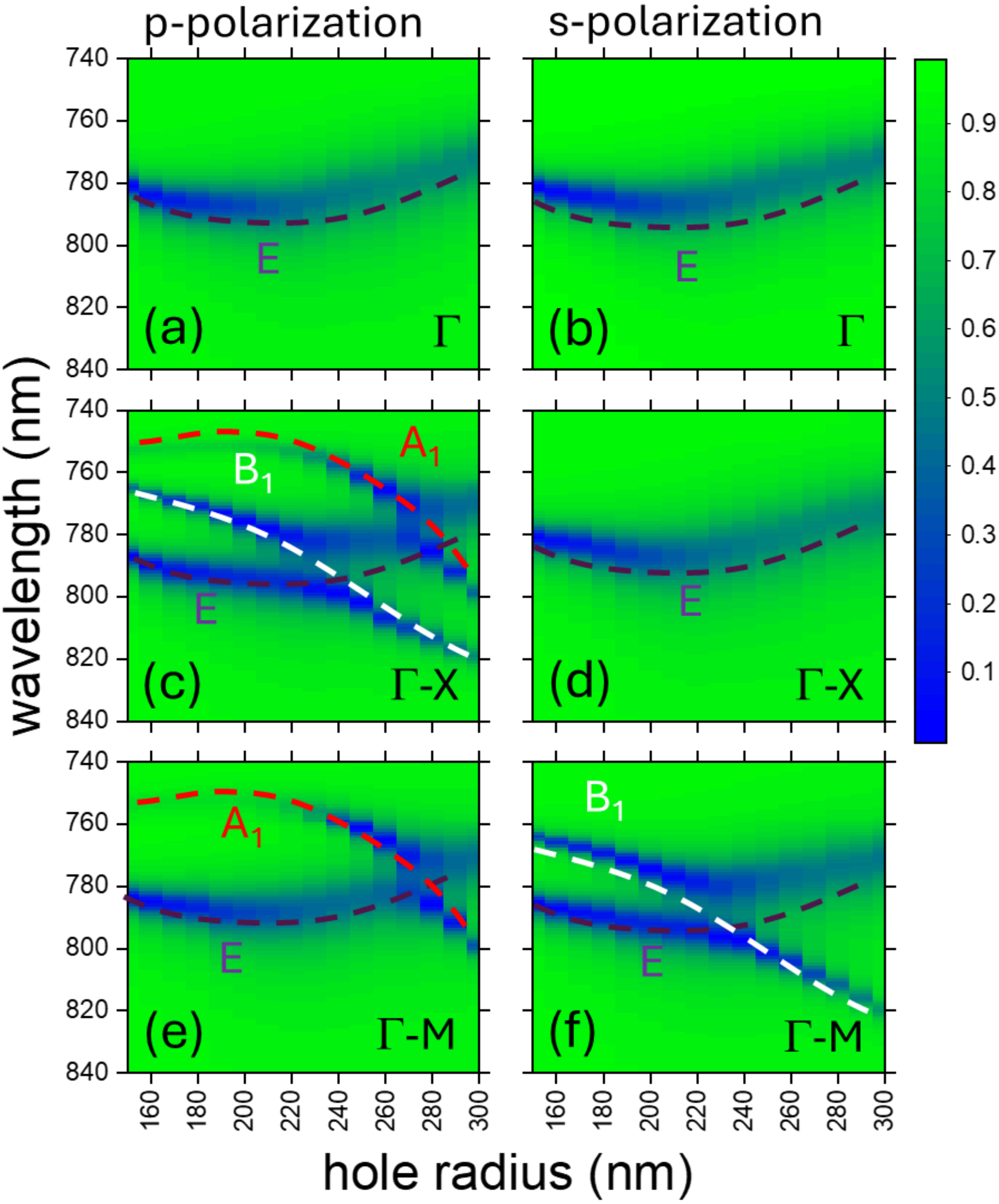

Fig. 2. The plots of the FDTD simulated (a) p- and (b) s-polarized specular (total) reflectivity mappings taken at the Γ point, where θ = 0°, as a function of R. The plots of the FDTD simulated (c) p- and (d) s-polarized reflectivity mappings taken along the Γ-X direction at θ = 1° as a function of R. The plots of the FDTD simulated (e) p- and (f) s-polarized reflectivity mappings taken along the Γ-M direction at θ = 1° as a function of R. The color dash lines are for the visualization of the $A_1$, $B_1$, and E modes.

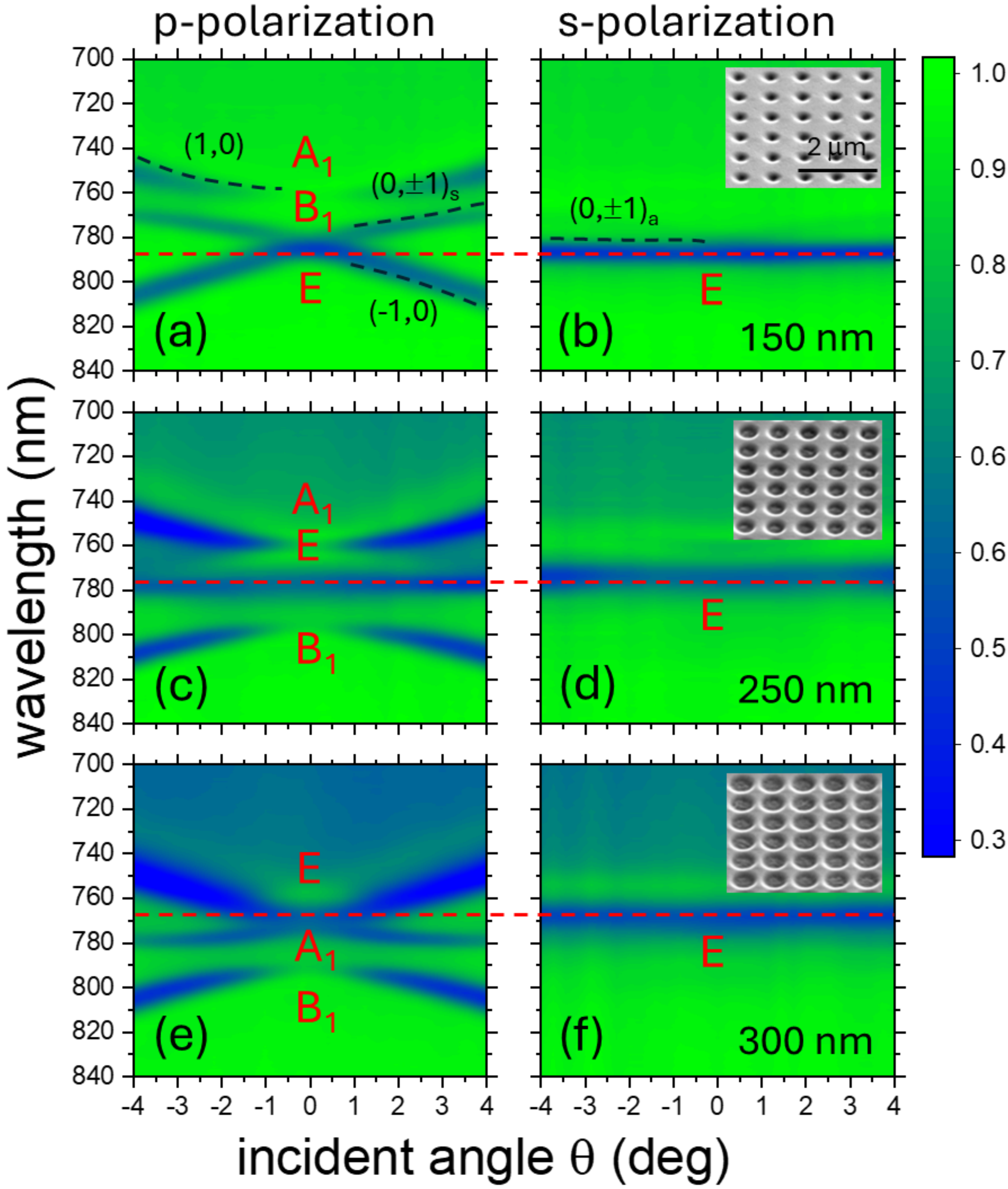

Fig. 3. The p-polarized angle-resolved specular reflectivity mappings measured near the Γ point along the Γ-X direction for R = (a) 150, (c) 250 and (e) 300 nm. $A_1$, $B_1$ and E are labelled at the Γ point. The s-polarized angle-resolved specular reflectivity mappings measured along the Γ-X direction for R = (b) 150, (c) 250 and (d) 300 nm. E is labelled at the Γ point. The dispersion relations are sketched as dash lines. The red dash lines are displayed for E mode degeneracy. Insets: the corresponding SEM images.

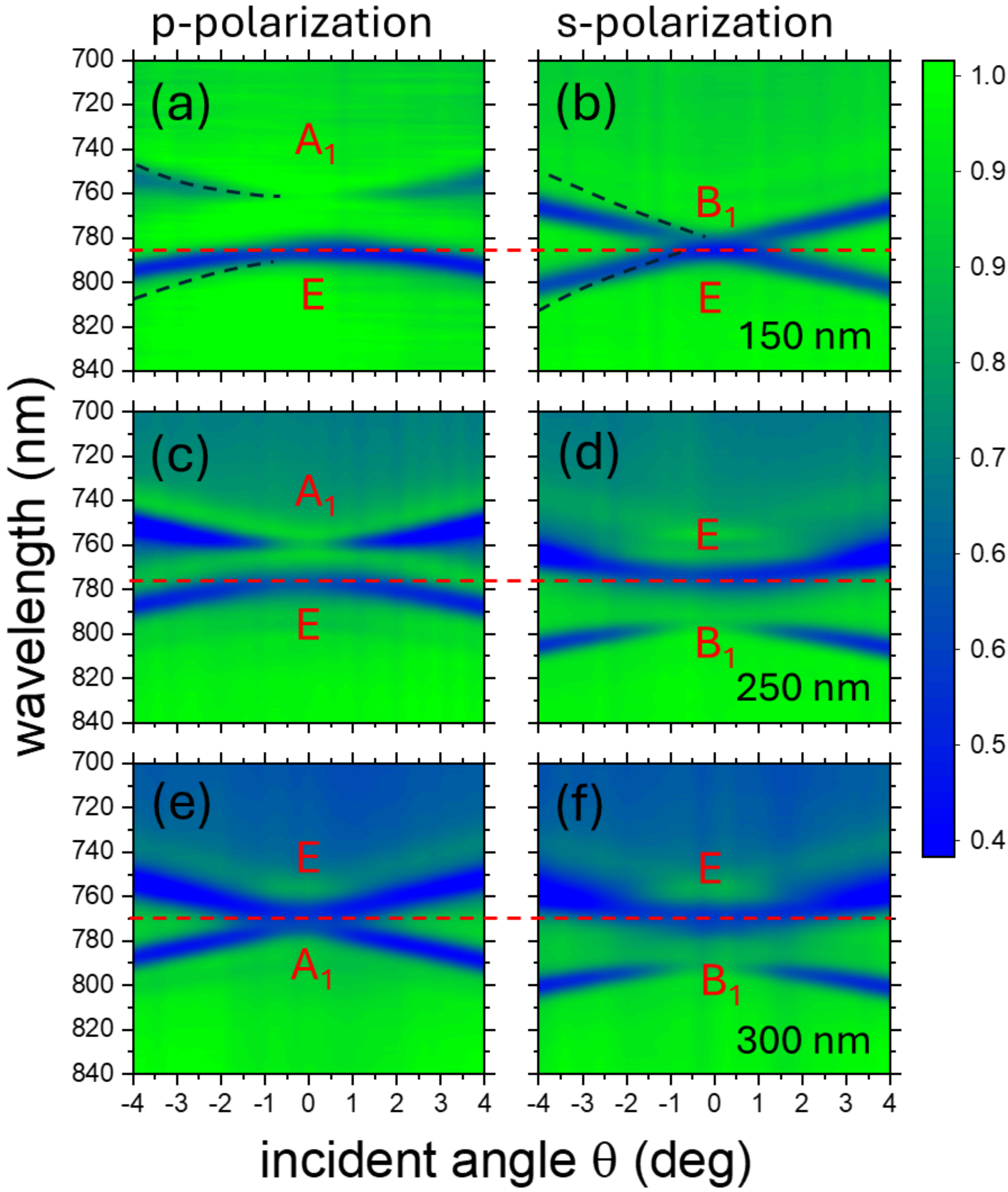

Fig. 4. The p-polarized angle-resolved specular reflectivity mappings measured near the Γ point along the Γ-M direction for R = (a) 150, (c) 250 and (e) 300 nm. $A_1$ and E are labelled at the Γ point. The s-polarized angle-resolved specular reflectivity mappings measured along the Γ-M direction for R = (b) 150, (d) 250 and (f) 300 nm. $B_1$ and E are labelled at the Γ point. The dispersion relations are sketched as dash lines. The red dash lines are displayed for E mode degeneracy.

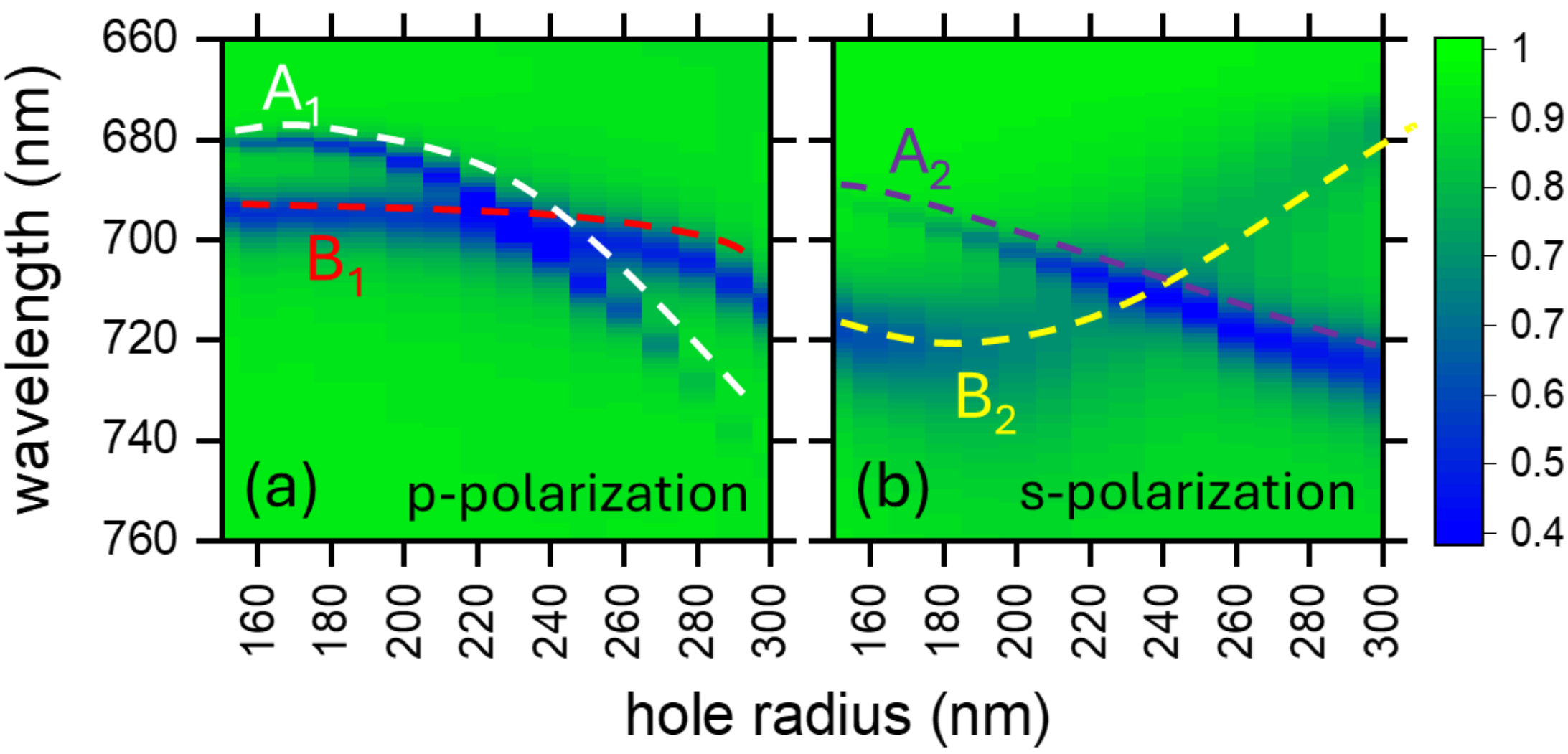

Fig. 5. The plots of the FDTD simulated (a) p- and (b) s-polarized total reflectivity mappings taken at the X point as a function of R. The color dash lines are for the visualization of the $A_1$, $B_1$, $A_2$ and $B_2$ modes.

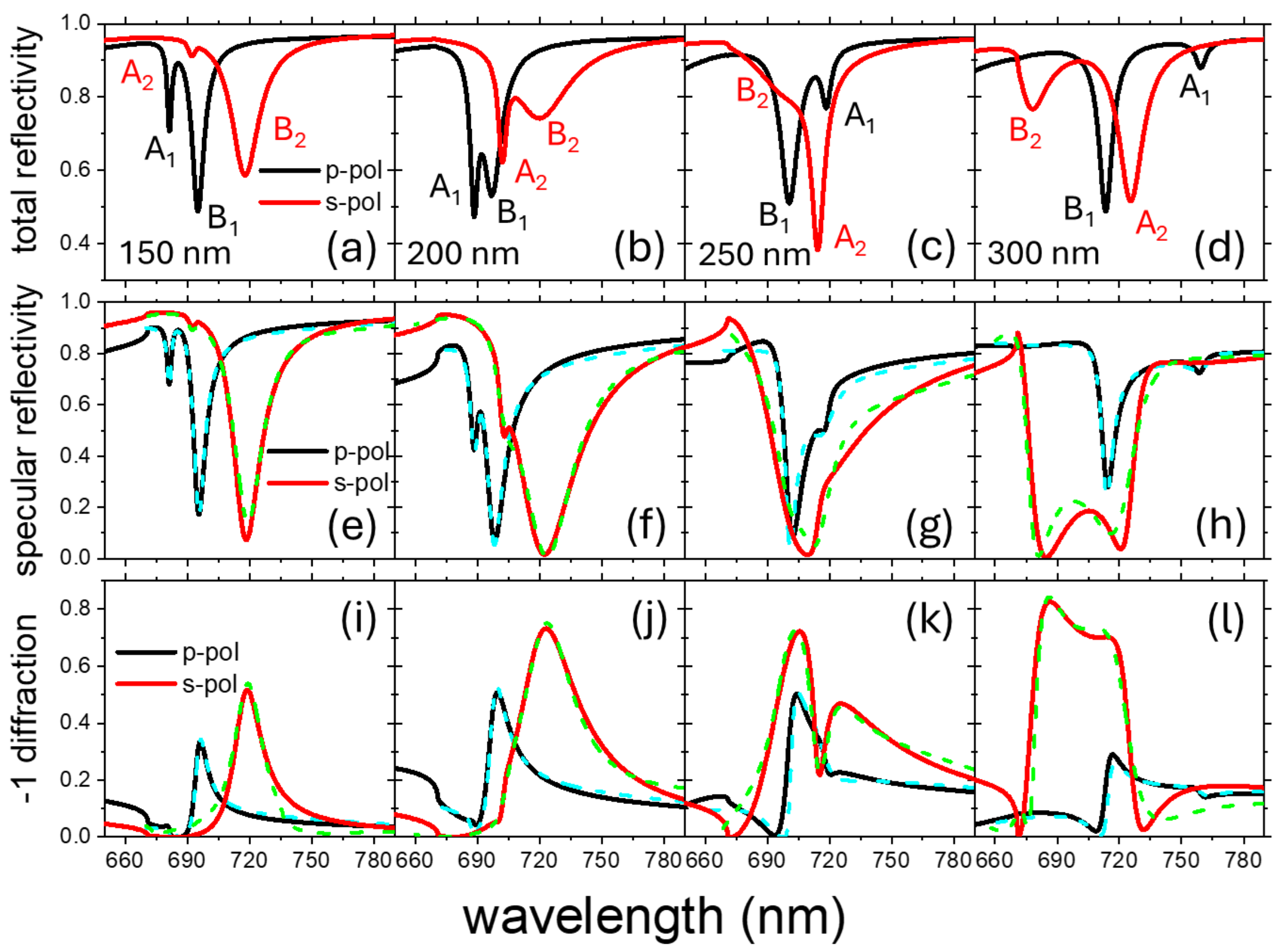

Fig. 6. The (black) p- and (red) s-polarized total reflectivity spectra taken at the X point for R = (a) 150, (b) 200, (c) 250, and (d) 300 nm. The (black) p- and (red) s-polarized specular reflectivity spectra taken at the X point for R = (e) 150, (f) 200, (g) 250, and (h) 300 nm. The best fits are displayed as dash lines. The (black) p- and (red) s-polarized $-1^{st}$ diffraction order spectra taken at the X point for R = (i) 150, (j) 200, (k) 250, and (l) 300 nm. The best fits are displayed as dash lines.

Table 1. The best fitted parameters for R = 150, 200, 250 and 300 nm

| p-pol | 150 nm | 200 nm | 250 nm | 300 nm |
|---|---|---|---|---|
| $\delta$ (meV) | -0.30 - 0.10*i* | -1.66 - 0.48*i* | -5.56 + 1.67*i* | -3.2 + 1.01*i* |
| $\tau$ (meV) | -4.10 + 1.13*i* | -7.37 + 2.98*i* | -2.10 | -0.28 - 0.32*i* |
| $\tilde{\omega}_-$ (eV) | 1.771 + 0.0099*i* | 1.770 + 0.011i | 1.73 + 0.0177*i* | 1.63 + 0.0114*i* |
| $\tilde{\omega}_+$ (eV) | 1.8198 + 0.0025*i* | 1.8011 + 0.0064*i* | 1.772 + 0.0084*i* | 1.7383 + 0.0081*i* |
| *d* | -1 | -1 | 1 | 1 |
| s-pol | 150 nm | 200 nm | 250 nm | 300 nm |
| $\mu$ (meV) | -0.10 | -3.08 + 0.6*i* | -12.82 - 1.36*i* | -17.40 + 6.47*i* |
| $\varepsilon$ (meV) | -12.47 + 0.89*i* | -29.85 + 5.08*i* | -19.48 + 4.04*i* | -14.23 - 7.35*i* |
| $\tilde{\omega}_-$ (eV) | 1.725 + 0.0205*i* | 1.7172 + 0.035*i* | 1.737 + 0.021*i* | 1.716 + 0.0305*i* |
| $\tilde{\omega}_+$ (eV) | 1.792 + 0.006*i* | 1.766 + 0.016*i* | 1.771 + 0.026*i* | 1.833 + 0.017*i* |
| *d* | -1 | -1 | 1 | 1 |

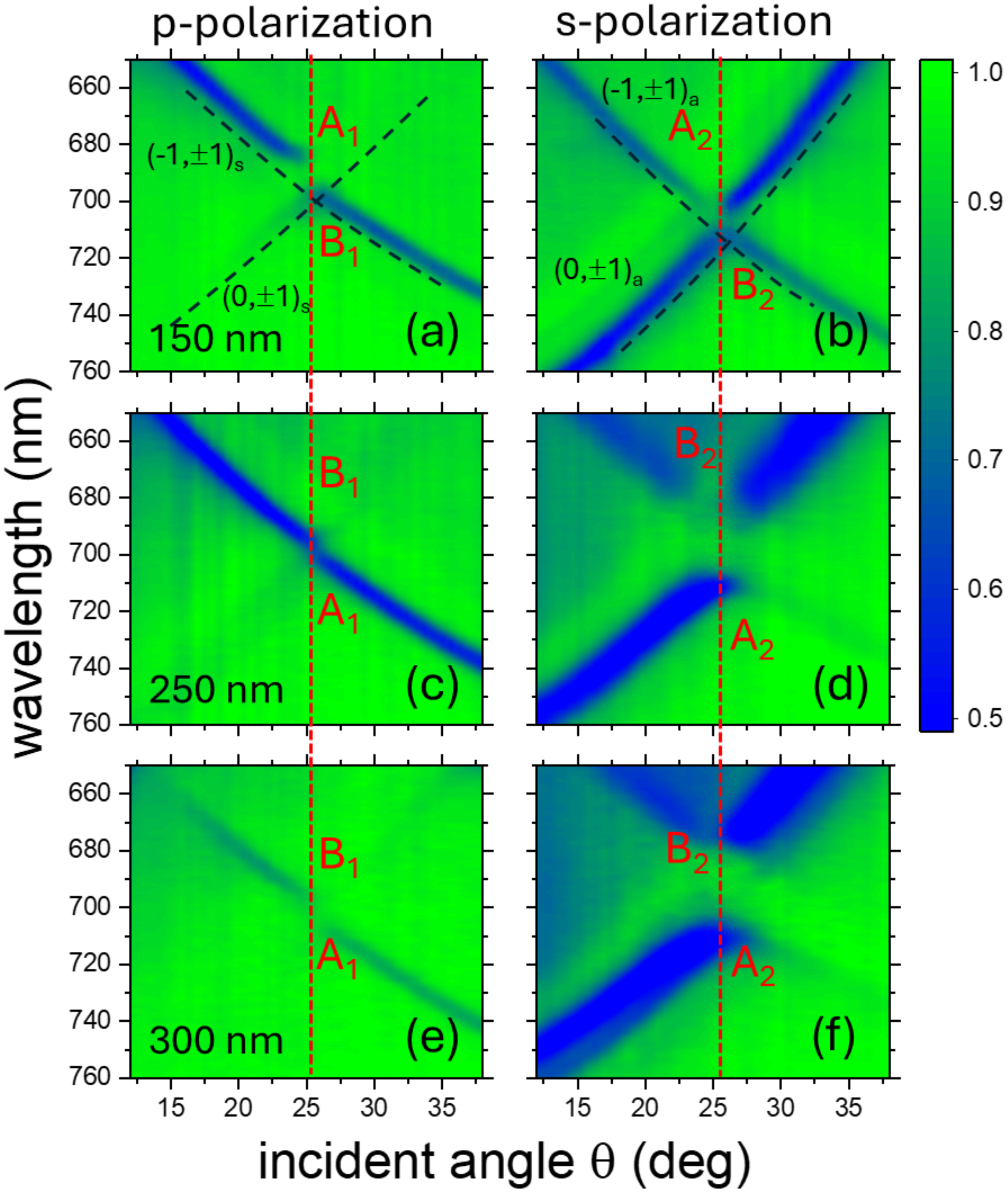

Fig. 7. The p-polarized angle-resolved specular reflectivity mappings measured near the X point along the Γ-X direction for R = (a) 150, (c) 250 and (e) 300 nm. $A_1$ and $B_1$ are labelled at the X point. (a) The dispersion relations are displayed as dash lines. The s-polarized angle-resolved specular reflectivity mappings measured near the X point along the Γ-X direction for R = (b) 150, (d) 250 and (f) 300 nm. $A_2$ and $B_2$ are labelled at the X point. (e) The dispersion relations are displayed as dash lines.

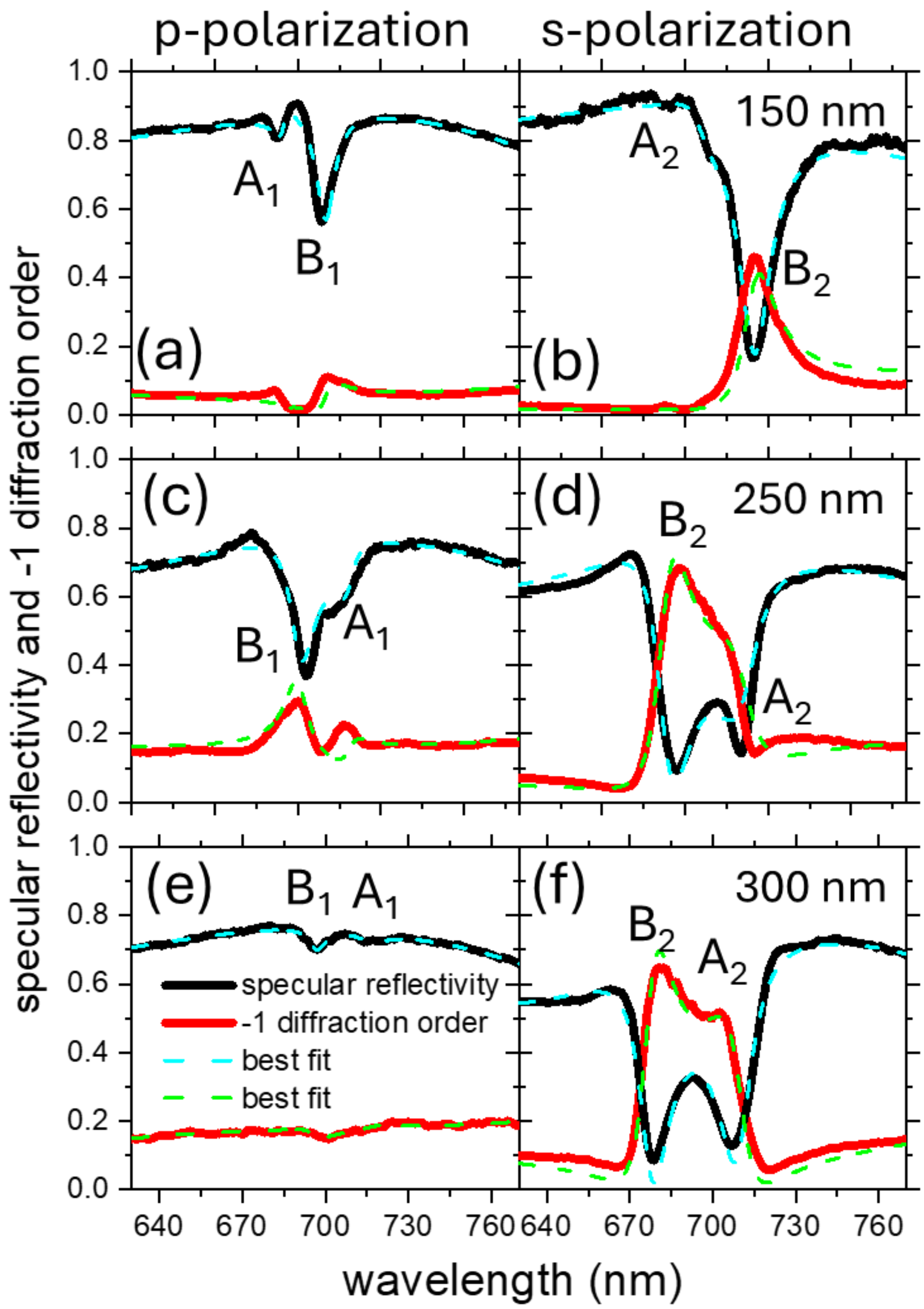

Fig. 8. The p-polarized specular reflectivity and -1st diffraction order spectra measured at the X point for R = (a) 150, (c) 250, and (e) 300 nm. The best fits are displayed as dash lines. The s-polarized specular reflectivity and -1st diffraction order spectra measured at the X point for R = (b) 150, (d) 250, and (f) 300 nm. The best fits are displayed as dash lines.

Table 2. The best fitted parameters for R = 150, 250 and 300 nm

| p-pol | 150 nm | 250 nm | 300 nm |
|---|---|---|---|
| $\delta$ (meV) | -0.20$i$ | -2.91 + 1.06$i$ | -0.30 + 0.04$i$ |
| $\tau$ (meV) | -1.9 - 0.02$i$ | -0.55 - 0.58$i$ | -0.29 + 0.07$i$ |
| $\tilde{\omega}_-$ (eV) | 1.771 + 0.01$i$ | 1.75 + 0.01$i$ | 1.73 + 0.0173$i$ |
| $\tilde{\omega}_+$ (eV) | 1.81 + 0.0044$i$ | 1.8 + 0.0137$i$ | 1.78 + 0.0095$i$ |
| $d$ | -1 | 1 | 1 |
| s-pol | 150 nm | 250 nm | 300 nm |
| $\mu$ (meV) | -8.4 + 1.1$i$ | -6.6 – 2.1$i$ | -9.9 – 1.4$i$ |
| $\varepsilon$ (meV) | -0.1 - 0.1$i$ | -20.2 + 3.8$i$ | -11.1 + 1.4$i$ |
| $\tilde{\omega}_-$ (eV) | 1.735 + 0.017$i$ | 1.743 + 0.023$i$ | 1.747 + 0.0184$i$ |
| $\tilde{\omega}_+$ (eV) | 1.77 + 0.0044$i$ | 1.815 + 0.019$i$ | 1.831 + 0.0173$i$ |
| $d$ | -1 | 1 | 1 |

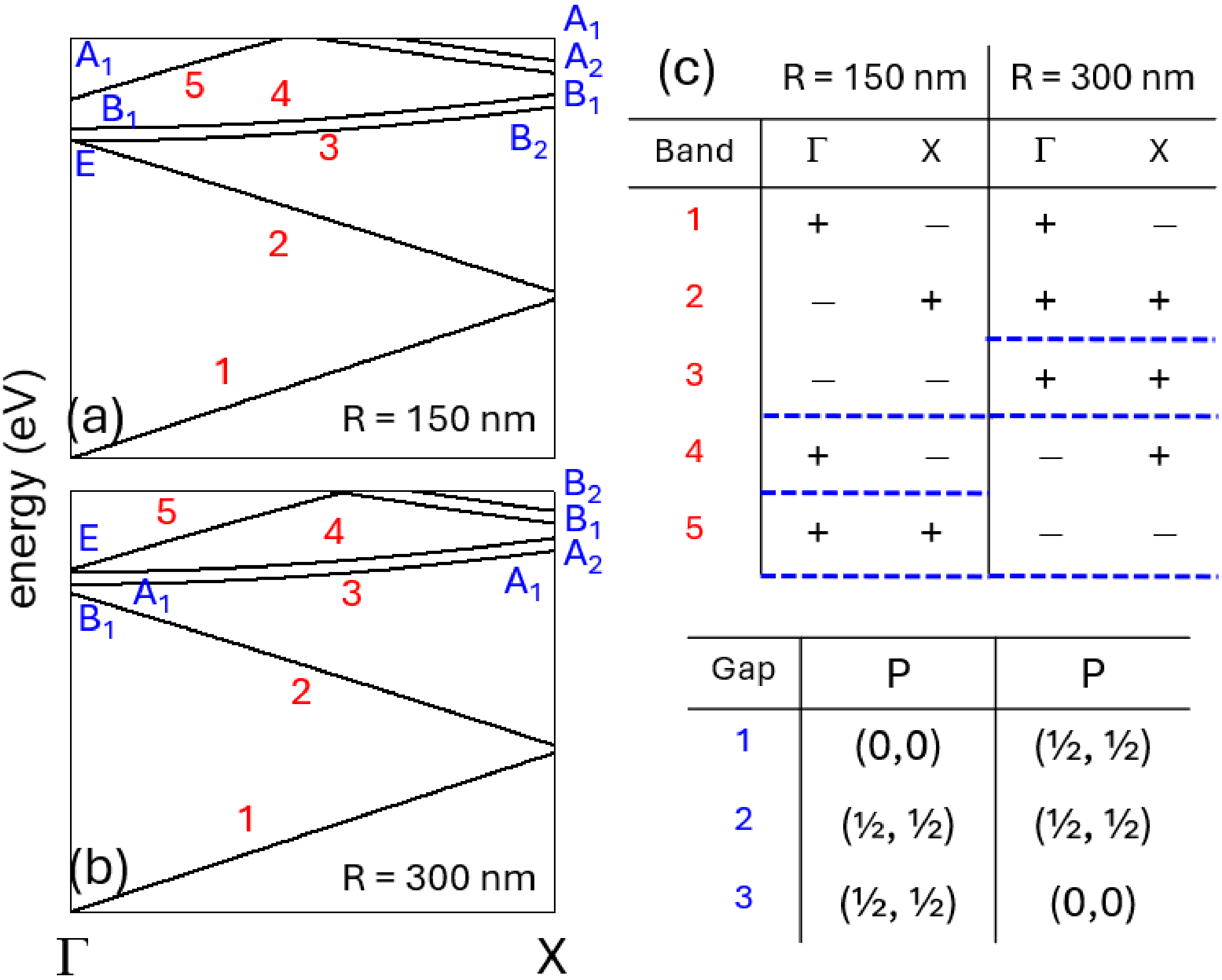

| | R = 150 nm | | R = 300 nm | |
|---|---|---|---|---|
| Band | Γ | X | Γ | X |
| 1 | + | − | + | − |
| 2 | − | + | + | + |
| 3 | − | − | + | + |
| 4 | + | − | − | + |
| 5 | + | + | − | − |

| Gap | P | P |
|---|---|---|
| 1 | (0,0) | (½, ½) |
| 2 | (½, ½) | (½, ½) |
| 3 | (½, ½) | (0,0) |

Fig. 9. The band structures of R = (a) 150 and (b) 300 nm PmCs are sketched. The energy bands are number and the modes at HSPs are labeled. (c) Top: the even (+) and odd (−) parities are tabulated. The dash lines indicate the band gaps. Bottom: the corresponding 2D polarizations P of the gaps.

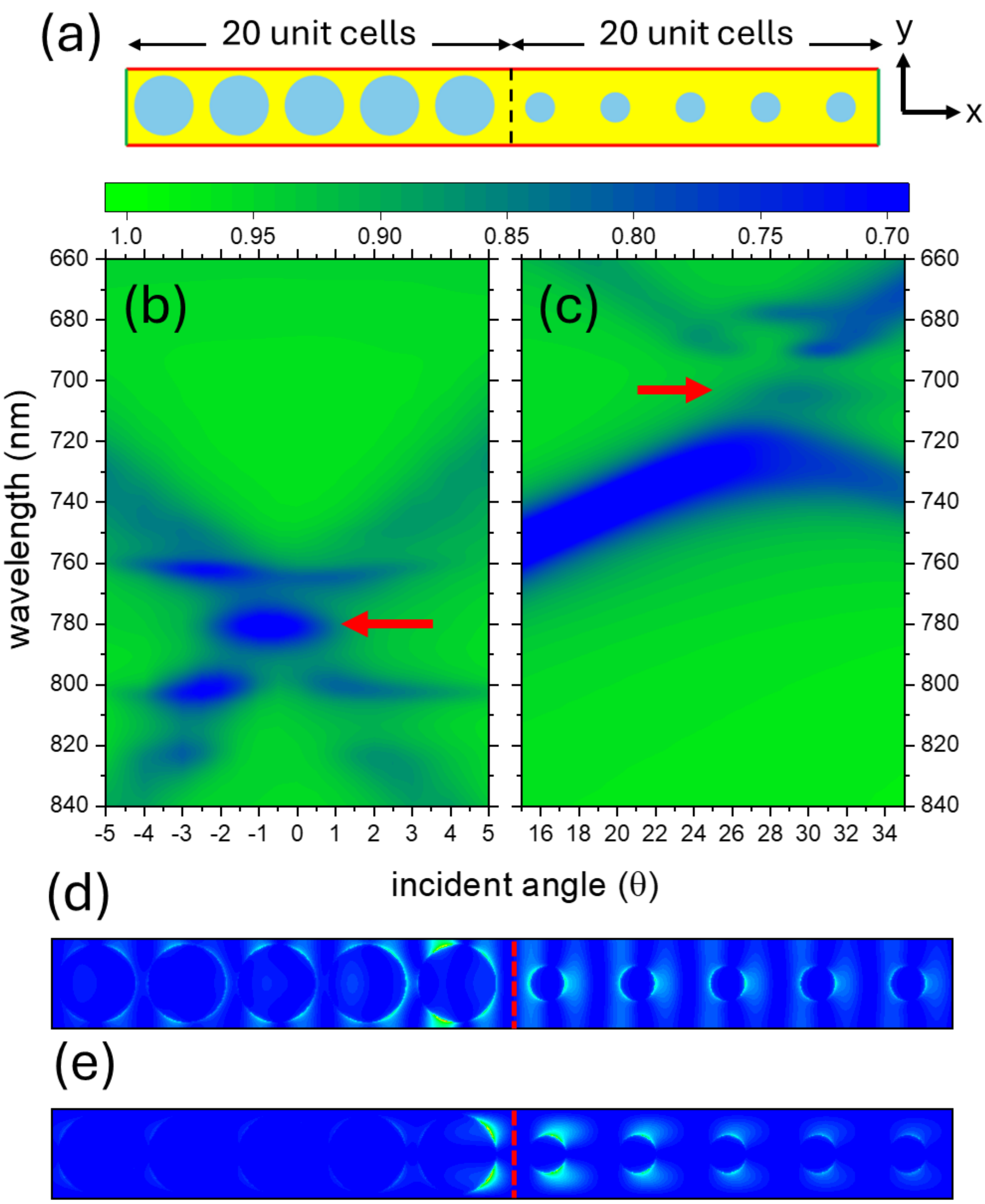

Fig. 10. (a) The schematic of the FDTD supercell for simulating the interface states. The angular reflectivity mappings of the supercell calculated at (b) the Γ point under p-polarization and (c) the X point under s-polarization. The arrows indicate the position of the edge states at 778 and 707 nm. The corresponding electric near-field patterns close to the interface, inducted by dash line, taken at (d) 778 and (e) 707 nm.

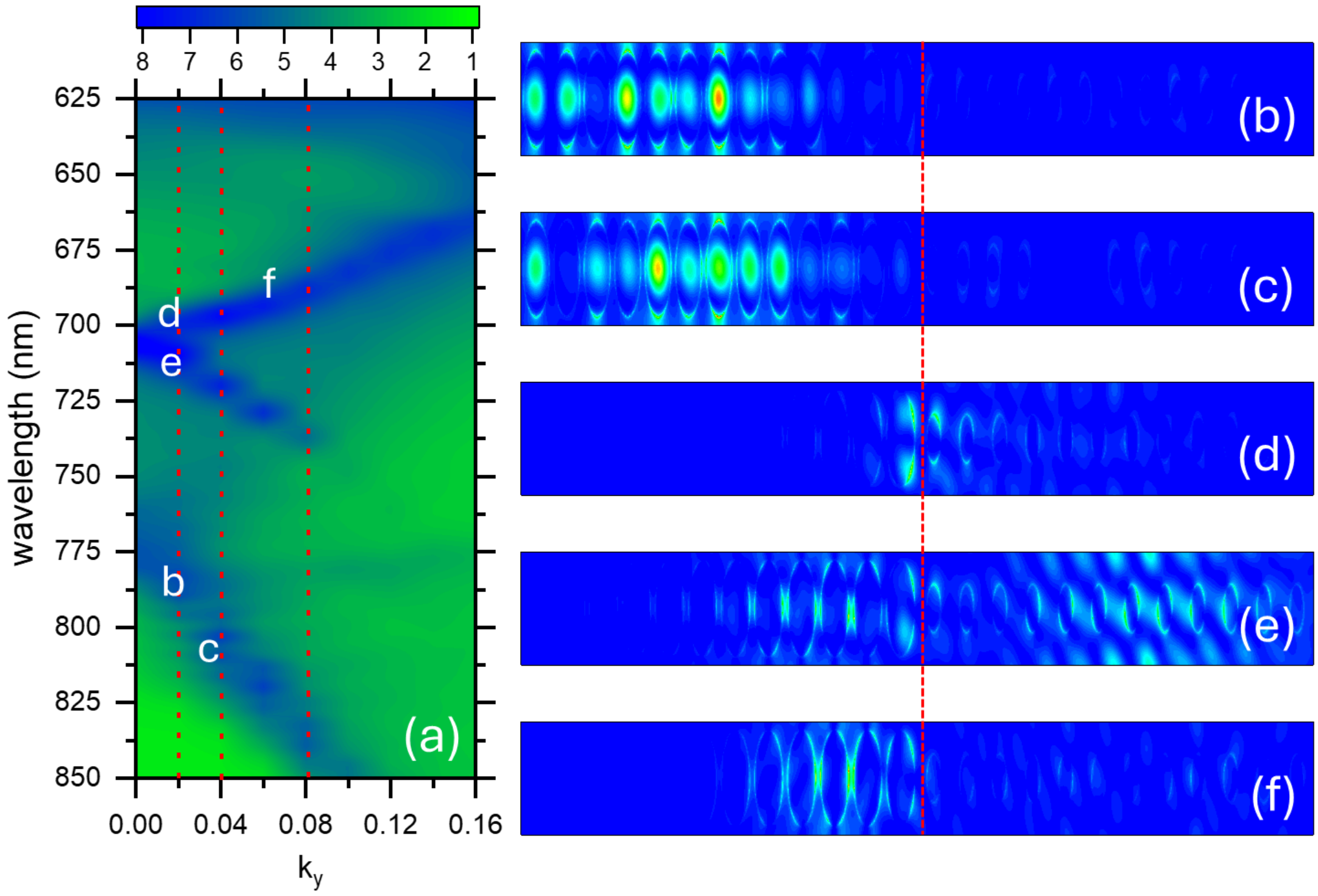

Fig. 11. (a) The $k_y$-dependent radiation spectral mapping of the supercell. The dash lines indicate $k_y$ = 0.02, 0.04 and 0.08. (b) – (f) The corresponding near-field patterns at different ($\lambda$,$k_y$) labeled in the mapping. The aspect ratios of the supercells are distorted for ease of visualization. The dash line shows the interface.

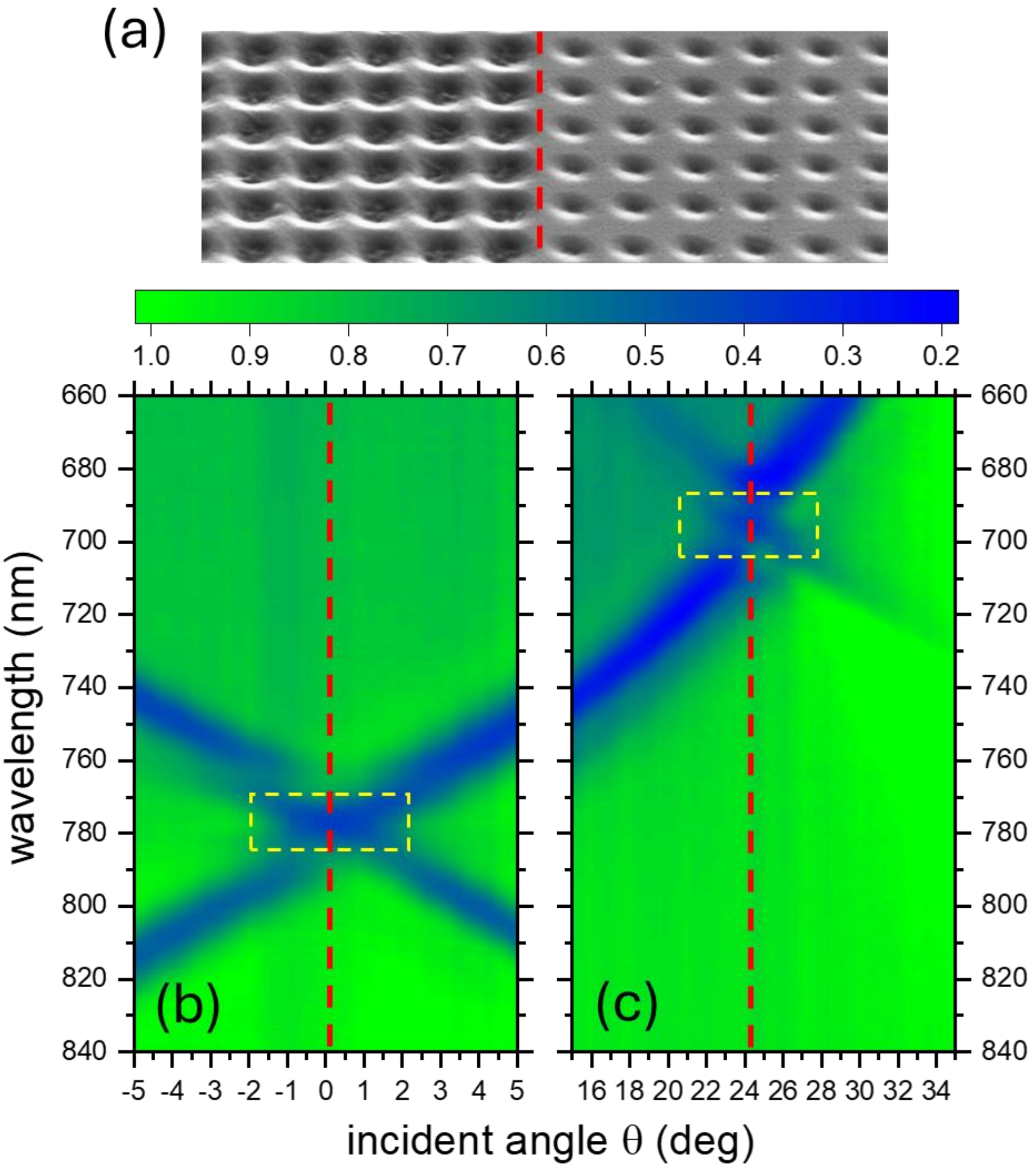

Fig. 12. (a) The SEM image of the heterojunction. The dash line is the location of the interface. The angular reflectivity spectral mappings taken at (a) the Γ point under p-polarization and (b) the X point under s-polarization. The dash lines are the Γ and X points and the dash boxes indicate the edge states.

# Supplementary Information

## Study of polarization-dependent band inversion and its associated edge state from two-dimensional square lattice plasmonic crystals

T.H. Chan, Y.H. Guan, C. Liu, and H.C. Ong

Department of Physics, The Chinese University of Hong Kong, Shatin, Hong Kong, People's Republic of China

### A. CMT formulation at the Γ point

As shown in Fig. S1(a), at the Γ point, four (±1,0) and (0,±1) SPPs propagate in the ±x and ±y directions. The CMT equation, which carries the general form of $\frac{dA}{dt} = iHA + KS_+$, can be expressed as:

$$\frac{d}{dt}\begin{bmatrix} a_1 \\ a_2 \\ a_3 \\ a_4 \end{bmatrix} = i\begin{bmatrix} \tilde{\omega}_o & \alpha & \beta & \beta \\ \alpha & \tilde{\omega}_o & \beta & \beta \\ \beta & \beta & \tilde{\omega}_o & \alpha \\ \beta & \beta & \alpha & \tilde{\omega}_o \end{bmatrix}\begin{bmatrix} a_1 \\ a_2 \\ a_3 \\ a_4 \end{bmatrix} + \begin{bmatrix} \kappa & 0 \\ -\kappa & 0 \\ 0 & \kappa \\ 0 & -\kappa \end{bmatrix}\begin{bmatrix} s_{p+} \\ s_{s+} \end{bmatrix}, \tag{S1}$$

where A is a column vector with the mode amplitudes $a_{1\text{-}4}$, H is a symmetric matrix which contains the degenerate complex angular frequency $\tilde{\omega}_o = \omega_o + i(\Gamma_a + \Gamma_r)/2$, where $\omega_o$ is the resonant frequency and $\Gamma_a$ and $\Gamma_r$ are the absorption and radiative decay rates, for $a_{1\text{-}4}$ and two complex interaction constants α and β. Because the Γ point possesses $C_{4v}$ symmetry, the complex interaction constant is defined as α for the modes propagating in the same direction but β for the modes propagating in the perpendicular direction. They are complex provided both near- and far-field interactions are present. In addition, for the in-coupling matrix K and the incident power $S_+$, each mode supports only one diffraction order with an in-coupling constant defined as κ, and the incident power amplitudes are defined as $s_{p+}$ and $s_{s+}$ for p- and s-polarizations in positive x and y directions. As shown in Fig. S1(b), the sign of κ depends on the relative orientation between the incident polarization and the transverse polarization of the SPP, which lies in the plane defined by $\vec{k}_{SPP}$. If they lie in the same direction, κ is positive, but it is negative when opposite. Four diffraction orders will interfere and yield a specular (total) reflection.

Once Eq. (S1) is formulated, we then solve the eigenvalues and eigenvectors by performing diagonalization. By using transformation matrix T such that $T\frac{dA}{dt} = iTHT^{-1}TA + TKS_+$, we have $\frac{d\tilde{A}}{dt} = i\tilde{H}\tilde{A} + \tilde{K}S_+$:

$$\frac{d}{dt}\begin{bmatrix} \tilde{a}_1 \\ \tilde{a}_2 \\ \tilde{a}_3 \\ \tilde{a}_4 \end{bmatrix} = i\begin{bmatrix} \tilde{\omega}_o - \alpha & 0 & 0 & 0 \\ 0 & \tilde{\omega}_o - \alpha & 0 & 0 \\ 0 & 0 & \tilde{\omega}_o + \alpha - 2\beta & 0 \\ 0 & 0 & 0 & \tilde{\omega}_o + \alpha + 2\beta \end{bmatrix}\begin{bmatrix} \tilde{a}_1 \\ \tilde{a}_2 \\ \tilde{a}_3 \\ \tilde{a}_4 \end{bmatrix} - \sqrt{2}\begin{bmatrix} 0 & \kappa \\ \kappa & 0 \\ 0 & 0 \\ 0 & 0 \end{bmatrix}\begin{bmatrix} s_{p+} \\ s_{s+} \end{bmatrix}, \quad \text{(S2)}$$

where $T = \frac{1}{2}\begin{bmatrix} 0 & 0 & -\sqrt{2} & \sqrt{2} \\ -\sqrt{2} & \sqrt{2} & 0 & 0 \\ -1 & -1 & 1 & 1 \\ 1 & 1 & 1 & 1 \end{bmatrix}$ and $T^{-1} = T^{T} = \frac{1}{2}\begin{bmatrix} 0 & -\sqrt{2} & -1 & 1 \\ 0 & \sqrt{2} & -1 & 1 \\ -\sqrt{2} & 0 & 1 & 1 \\ \sqrt{2} & 0 & 1 & 1 \end{bmatrix}$. From Eq. (S2), we see from $\tilde{H}$ that $\tilde{a}_{1,2}$ modes are degenerate with $\tilde{\omega}_{1,2} = \tilde{\omega}_o - \alpha$ but $\tilde{a}_{3,4}$ modes, $\tilde{\omega}_{3,4} = \tilde{\omega}_o + \alpha \mp 2\beta$, are nondegenerate. Assuming $a_{1\text{-}4}$ have the spatial form of $e^{i\vec{k}_{SPP}\cdot\vec{r}}$, or $e^{\pm i\frac{2\pi}{P}x}$ and $e^{\pm i\frac{2\pi}{P}y}$, their electric field magnitudes in the z direction, $E_z(x,y)$, can be roughly approximated as:

$$\begin{bmatrix} \tilde{a}_1 \\ \tilde{a}_2 \\ \tilde{a}_3 \\ \tilde{a}_4 \end{bmatrix} = T\begin{bmatrix} a_1 \\ a_2 \\ a_3 \\ a_4 \end{bmatrix} = \begin{bmatrix} \frac{-a_3 + a_4}{\sqrt{2}} \\ \frac{-a_1 + a_2}{\sqrt{2}} \\ \frac{-a_1 - a_2 + a_3 + a_4}{2} \\ \frac{a_1 + a_2 + a_3 + a_4}{2} \end{bmatrix} = \begin{bmatrix} -\sqrt{2}i\sin(2\pi y/P) \\ -\sqrt{2}i\sin(2\pi x/P) \\ -\cos(2\pi x/P) + \cos(2\pi y/P) \\ \cos(2\pi x/P) + \cos(2\pi y/P) \end{bmatrix}.$$

Their field patterns are illustrated in the following Fig. S1(c), and they are assigned as E, $B_1$ and $A_1$ for $\tilde{a}_{1,2}$, $\tilde{a}_3$, and $\tilde{a}_4$. In other words, $\tilde{a}_{1,2}$ (E) exhibit p character and $\tilde{a}_3$ ($B_1$) and $\tilde{a}_4$ ($A_1$) display $d_{xy}$ and s characters.

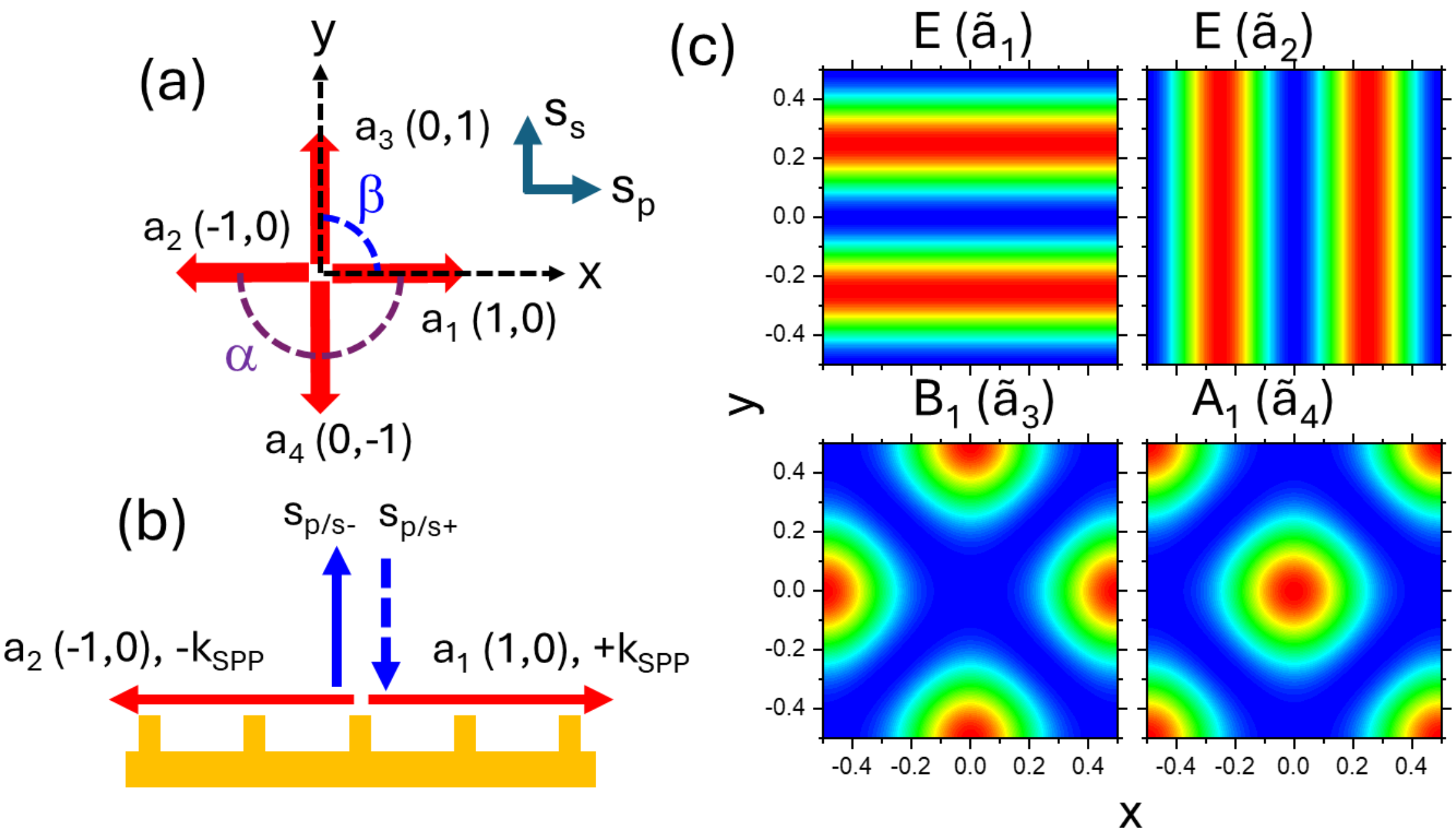

Fig. S1. (a) The schematic plane-view of how four (±1,0) and (0,±1) SPPs couple at the Γ point for CMT formulation and the coupling constants are defined as α and β. (b) The cross-section view shows the incoming $s_+$ and the outgoing specular (total) reflectivity $s_-$ channels at the Γ point. Each channel carries p- and s-polarizations. (c) The field patterns of $\tilde{a}_{1-4}$ and are labelled as E, $B_1$, and $A_1$.

## B. CMT formulation at the X point

As shown in S2(a), at the X point, four degenerate (0,±1) and (-1,±1) SPPs couple together and the CMT equation can be expressed as:

$$\frac{d}{dt}\begin{bmatrix} a_1 \\ a_2 \\ a_3 \\ a_4 \end{bmatrix} = i\begin{bmatrix} \tilde{\omega}_o & \sigma & \gamma & \eta \\ \sigma & \tilde{\omega}_o & \eta & \gamma \\ \gamma & \eta & \tilde{\omega}_o & \sigma \\ \eta & \gamma & \sigma & \tilde{\omega}_o \end{bmatrix}\begin{bmatrix} a_1 \\ a_2 \\ a_3 \\ a_4 \end{bmatrix} + \begin{bmatrix} -\kappa_{pL} & \kappa_{sL} & -\kappa_{pR} & \kappa_{sR} \\ \kappa_{pR} & -\kappa_{sR} & \kappa_{pL} & -\kappa_{sL} \\ -\kappa_{pL} & -\kappa_{sL} & -\kappa_{pR} & -\kappa_{sR} \\ \kappa_{pR} & \kappa_{sR} & \kappa_{pL} & \kappa_{sL} \end{bmatrix}\begin{bmatrix} s_{pL+} \\ s_{sL+} \\ s_{pR+} \\ s_{sR+} \end{bmatrix}, \tag{S3}$$

where the complex coupling constants are η, γ, and σ for the modes oriented relatively to each other by 60°, 120°, and 180°, respectively. As the X point follows $C_{2v}$ symmetry in which only two right- and left-hand diffractions are supported, the in-coupling matrix K has right and left input channels and each channel carries p- and s-polarizations, as shown in Fig. S2(b). The first p/s and second R/L subscripts for κ and $s_+$ in K and $S_+$ in Eq. (S3) define the p/s polarizations and the right/left incidences.

After diagonalization, Eq. (S3) is transformed as:

$$\frac{d}{dt}\begin{bmatrix} \tilde{a}_1 \\ \tilde{a}_2 \\ \tilde{a}_3 \\ \tilde{a}_4 \end{bmatrix} = i\begin{bmatrix} \tilde{\omega}_o+\sigma-\gamma-\eta & 0 & 0 & 0 \\ 0 & \tilde{\omega}_o-\sigma+\gamma-\eta & 0 & 0 \\ 0 & 0 & \tilde{\omega}_o-\sigma-\gamma+\eta & 0 \\ 0 & 0 & 0 & \tilde{\omega}_o+\sigma+\gamma+\eta \end{bmatrix}\begin{bmatrix} \tilde{a}_1 \\ \tilde{a}_2 \\ \tilde{a}_3 \\ \tilde{a}_4 \end{bmatrix} + TK\begin{bmatrix} s_{pL+} \\ s_{sL+} \\ s_{pR+} \\ s_{sR+} \end{bmatrix}, \tag{S4}$$

where the transformation matrices T and $T^{-1} = T^T$ are $\frac{1}{2}\begin{bmatrix} -1 & -1 & 1 & 1 \\ -1 & 1 & -1 & 1 \\ 1 & -1 & -1 & 1 \\ 1 & 1 & 1 & 1 \end{bmatrix}$ and $\frac{1}{2}\begin{bmatrix} -1 & -1 & 1 & 1 \\ -1 & 1 & -1 & 1 \\ 1 & -1 & -1 & 1 \\ 1 & 1 & 1 & 1 \end{bmatrix}$, respectively. We see all the coupled modes are nondegenerate. As a result, $TKS_+ = \begin{bmatrix} 0 & \kappa_{sR}-\kappa_{sL} & 0 & -(\kappa_{sR}-\kappa_{sL}) \\ \kappa_{pR}+\kappa_{pL} & 0 & \kappa_{pR}+\kappa_{pL} & 0 \\ 0 & \kappa_{sR}+\kappa_{sL} & 0 & \kappa_{sR}+\kappa_{sL} \\ \kappa_{pR}-\kappa_{pL} & 0 & -(\kappa_{pR}-\kappa_{pL}) & 0 \end{bmatrix}\begin{bmatrix} s_{pL+} \\ s_{sL+} \\ s_{pR+} \\ s_{sR+} \end{bmatrix}$, which is equal to

$$\begin{bmatrix} (\kappa_{sR}-\kappa_{sL})s_{sL+}-(\kappa_{sR}-\kappa_{sL})s_{sR+} \\ (\kappa_{pR}+\kappa_{pL})s_{pL+}+(\kappa_{pR}+\kappa_{pL})s_{pR+} \\ (\kappa_{sR}+\kappa_{sL})s_{sL+}+(\kappa_{sR}+\kappa_{sL})s_{sR+} \\ (\kappa_{pR}-\kappa_{pL})s_{pL+}-(\kappa_{pR}-\kappa_{pL})s_{pR+} \end{bmatrix}$$

, and

$$\begin{bmatrix} \tilde{a}_1 \\ \tilde{a}_2 \\ \tilde{a}_3 \\ \tilde{a}_4 \end{bmatrix} = \frac{1}{2}\begin{bmatrix} -a_1-a_2+a_3+a_4 \\ -a_1+a_2-a_3+a_4 \\ a_1-a_2-a_3+a_4 \\ a_1+a_2+a_3+a_4 \end{bmatrix} =$$

$$= 2\begin{bmatrix} -\sin(\pi x/P)\sin(2\pi y/P) \\ i\sin(\pi x/P)\cos(2\pi y/P) \\ i\cos(\pi x/P)\sin(2\pi y/P) \\ \cos(\pi x/P)\cos(2\pi y/P) \end{bmatrix}$$

, assuming $a_{1\text{-}4}$ have the spatial form of $e^{i\vec{k}_{SPP}\cdot\vec{r}}$ , or $e^{\pm i(x\pm 2y)(\pi/P)}$.

From Eq. (S4), we see $\tilde{a}_1 = \frac{(\kappa_{sR}-\kappa_{sL})(s_{sL+}-s_{sR+})}{i(\omega-\tilde{\omega}_1)}$ , $\tilde{a}_2 = \frac{(\kappa_{pR}+\kappa_{pL})(s_{pL+}+s_{pR+})}{i(\omega-\tilde{\omega}_2)}$ , $\tilde{a}_3 = \frac{(\kappa_{sR}+\kappa_{sL})(s_{sL+}+s_{sR+})}{i(\omega-\tilde{\omega}_3)}$ , and $\tilde{a}_4 = \frac{(\kappa_{pR}-\kappa_{pL})(s_{pL+}-s_{pR+})}{i(\omega-\tilde{\omega}_4)}$ . The corresponding field patterns are illustrated in Fig. S2(c) and $\tilde{a}_1$ and $\tilde{a}_3$ are assigned to be $A_2$ and $B_2$ where $\tilde{a}_2$ and $\tilde{a}_4$ are assigned to be $B_1$ and $A_1$.

We then formulate their right and left diffraction orders. We define the specular reflection and the -1st diffraction order as follows. As shown in Fig. S2(b), if the incidence is the right incoming channel ($s_{s/pR+}$), then the specular reflection and the -1st diffraction orders are the left ($s_{s/pL-}$) and right ($s_{s/pR-}$) outgoing channels. Therefore, by using conservation of energy and time reversal symmetry, the outgoing radiation channels are expressed as $S_- = CS_+ + K^T T^T \tilde{A}$ or

$$\begin{bmatrix} s_{pL-} \\ s_{sL-} \\ s_{pR-} \\ s_{sR-} \end{bmatrix} = C\begin{bmatrix} s_{pL+} \\ s_{sL+} \\ s_{pR+} \\ s_{sR+} \end{bmatrix} + K^T T^T \begin{bmatrix} \tilde{a}_1 \\ \tilde{a}_2 \\ \tilde{a}_3 \\ \tilde{a}_4 \end{bmatrix}, \tag{S5}$$

where C is the non-resonant scattering matrix and

$$K^T T^T = \frac{1}{2}\begin{bmatrix} -\kappa_{pL} & \kappa_{pR} & -\kappa_{pL} & \kappa_{pR} \\ \kappa_{sL} & -\kappa_{sR} & -\kappa_{sL} & \kappa_{sR} \\ -\kappa_{pR} & \kappa_{pL} & -\kappa_{pR} & \kappa_{pL} \\ \kappa_{sR} & -\kappa_{sL} & -\kappa_{sR} & \kappa_{sL} \end{bmatrix}\begin{bmatrix} -1 & -1 & 1 & 1 \\ -1 & 1 & -1 & 1 \\ 1 & -1 & -1 & 1 \\ 1 & 1 & 1 & 1 \end{bmatrix} =$$

$$\begin{bmatrix} 0 & \kappa_{pR}+\kappa_{pL} & 0 & \kappa_{pR}-\kappa_{pL} \\ \kappa_{sR}-\kappa_{sL} & 0 & \kappa_{sR}+\kappa_{sL} & 0 \\ 0 & \kappa_{pR}+\kappa_{pL} & 0 & -(\kappa_{pR}-\kappa_{pL}) \\ -(\kappa_{sR}-\kappa_{sL}) & 0 & \kappa_{sR}+\kappa_{sL} & 0 \end{bmatrix}$$

. As a result, Eq. (S5) becomes:

$$\begin{bmatrix} s_{pL-} \\ s_{sL-} \\ s_{pR-} \\ s_{sR-} \end{bmatrix} = C \begin{bmatrix} s_{pL+} \\ s_{sL+} \\ s_{pR+} \\ s_{sR+} \end{bmatrix} + \begin{bmatrix} 0 \\ \kappa_{sR} - \kappa_{sL} \\ 0 \\ -\left(\kappa_{sR} - \kappa_{sL}\right) \end{bmatrix} \tilde{a}_1 + \begin{bmatrix} \kappa_{pR} + \kappa_{pL} \\ 0 \\ \kappa_{pR} + \kappa_{pL} \\ 0 \end{bmatrix} \tilde{a}_2 + \begin{bmatrix} 0 \\ \kappa_{sR} + \kappa_{sL} \\ 0 \\ \kappa_{sR} + \kappa_{sL} \end{bmatrix} \tilde{a}_3 + \begin{bmatrix} \kappa_{pR} - \kappa_{pL} \\ 0 \\ -\left(\kappa_{pR} - \kappa_{pL}\right) \\ 0 \end{bmatrix} \tilde{a}_4 \text{ ,(S6)}$$

By taking only $s_{sR+} \neq 0$ under single right s-incidence, the diffraction orders are:

$$\begin{bmatrix} s_{pL-} \\ s_{sL-} \\ s_{pR-} \\ s_{sR-} \end{bmatrix} = C \begin{bmatrix} 0 \\ 0 \\ 0 \\ s_{sR+} \end{bmatrix} + \begin{bmatrix} 0 \\ -\left(\kappa_{sR} - \kappa_{sL}\right)^2 \\ 0 \\ \left(\kappa_{sR} - \kappa_{sL}\right)^2 \end{bmatrix} \frac{s_{sR+}}{i\left(\omega - \tilde{\omega}_1\right)} + \begin{bmatrix} 0 \\ \left(\kappa_{sR} + \kappa_{sL}\right)^2 \\ 0 \\ \left(\kappa_{sR} + \kappa_{sL}\right)^2 \end{bmatrix} \frac{s_{sR+}}{i\left(\omega - \tilde{\omega}_3\right)}, \quad \text{(S7)}$$

showing only two diffractions at $\tilde{\omega}_1$ ($A_2$) and $\tilde{\omega}_3$ ($B_2$) are present and they both are s-polarized. In addition, while the left (specular reflectivity) and right (-1 diffraction order) diffractions from $B_2$ are in phase, those from $A_2$ are $\pi$ out of phase. Likewise, for $s_{pR+} \neq 0$, we have

$$\begin{bmatrix} s_{pL-} \\ s_{sL-} \\ s_{pR-} \\ s_{sR-} \end{bmatrix} = C \begin{bmatrix} 0 \\ 0 \\ s_{pR+} \\ 0 \end{bmatrix} + \begin{bmatrix} \left(\kappa_{pR} + \kappa_{pL}\right)^2 \\ 0 \\ \left(\kappa_{pR} + \kappa_{pL}\right)^2 \\ 0 \end{bmatrix} \frac{s_{pR+}}{i\left(\omega - \tilde{\omega}_2\right)} + \begin{bmatrix} -\left(\kappa_{pR} - \kappa_{pL}\right)^2 \\ 0 \\ \left(\kappa_{pR} - \kappa_{pL}\right)^2 \\ 0 \end{bmatrix} \frac{s_{pR+}}{i\left(\omega - \tilde{\omega}_4\right)}, \quad \text{(S8)}$$

indicating only p-polarized diffractions at $\tilde{\omega}_2$ ($B_1$) and $\tilde{\omega}_4$ ($A_1$) exist. The left (specular reflectivity) and right (-1 diffraction order) diffractions are in phase for $B_1$ but $\pi$ out of phase for $A_1$.

Eq. (S7) and (S8) can be further simplified. We rewrite them for the Frensel $r_{p/s,0}$ and $r_{p/s,-1}$ coefficients as two superimposed Lorentzian profiles given as:

$$\begin{bmatrix} r_{p,0} \\ r_{p,-1} \end{bmatrix} = \begin{bmatrix} p_{0,b} \\ p_{-1,b} e^{i\frac{\pi}{2}} \end{bmatrix} + \begin{bmatrix} \dfrac{\left(\kappa_{pR} + \kappa_{pL}\right)^2}{i\left(\omega - \tilde{\omega}_2\right)} - \dfrac{\left(\kappa_{pR} - \kappa_{pL}\right)^2}{i\left(\omega - \tilde{\omega}_4\right)} \\ \dfrac{\left(\kappa_{pR} + \kappa_{pL}\right)^2}{i\left(\omega - \tilde{\omega}_2\right)} + \dfrac{\left(\kappa_{pR} - \kappa_{pL}\right)^2}{i\left(\omega - \tilde{\omega}_4\right)} \end{bmatrix} = \begin{bmatrix} p_{0,b} \\ ip_{-1,b} \end{bmatrix} + \begin{bmatrix} \dfrac{\delta}{i\left(\omega - \tilde{\omega}_2\right)} - \dfrac{\gamma}{i\left(\omega - \tilde{\omega}_4\right)} \\ \dfrac{\delta}{i\left(\omega - \tilde{\omega}_2\right)} + \dfrac{\gamma}{i\left(\omega - \tilde{\omega}_4\right)} \end{bmatrix} \text{ or}$$

$$\begin{bmatrix} r_{p,0} \\ r_{p,-1} \end{bmatrix} = \begin{bmatrix} p_{0,b} \\ ip_{-1,b} \end{bmatrix} + \begin{bmatrix} \dfrac{\delta}{i\left(\omega - \tilde{\omega}_+\right)} - \dfrac{\gamma}{i\left(\omega - \tilde{\omega}_-\right)} \\ d\left( \dfrac{\delta}{i\left(\omega - \tilde{\omega}_+\right)} + \dfrac{\gamma}{i\left(\omega - \tilde{\omega}_-\right)} \right) \end{bmatrix}, \quad \text{(S9)}$$

and

$$\begin{bmatrix} r_{s,0} \\ r_{s,-1} \end{bmatrix} = \begin{bmatrix} s_{0,b} \\ s_{-1,b} e^{i\frac{\pi}{2}} \end{bmatrix} + \begin{bmatrix} \frac{-(\kappa_{sR} - \kappa_{sL})^2}{i(\omega - \tilde{\omega}_1)} + \frac{(\kappa_{sR} + \kappa_{sL})^2}{i(\omega - \tilde{\omega}_3)} \\ \frac{(\kappa_{sR} - \kappa_{sL})^2}{i(\omega - \tilde{\omega}_1)} + \frac{(\kappa_{sR} + \kappa_{sL})^2}{i(\omega - \tilde{\omega}_3)} \end{bmatrix} \text{ or}$$

$$\begin{bmatrix} r_{s,0} \\ r_{s,-1} \end{bmatrix} = \begin{bmatrix} s_{0,b} \\ is_{-1,b} \end{bmatrix} + \begin{bmatrix} \frac{\mu}{i(\omega - \tilde{\omega}_+)} + \frac{\varepsilon}{i(\omega - \tilde{\omega}_-)} \\ d\left( \frac{\mu}{i(\omega - \tilde{\omega}_+)} - \frac{\varepsilon}{i(\omega - \tilde{\omega}_-)} \right) \end{bmatrix}, \tag{S10}$$

where $p/s_{0,b}$ and $p/s_{-1,b}$ are the nonresonant backgrounds, $\tilde{\omega}_-$ and $\tilde{\omega}_+$ are the complex angular frequencies for the coupled modes, and $\delta$, $\gamma$, $\mu$, and $\varepsilon$ are the constants. The phase differences between $p/s_{0,b}$ and $p/s_{-1,b}$ are known to be $-\pi/2$ [1,2]. The symmetry factor d is either 1 or -1, which determines whether $\tilde{\omega}_+$ is $A_1$ or $B_1$ under p-polarization and $A_2$ or $B_2$ under s-polarization. We can fit the specular reflectivity and -1 diffraction order spectra by using $|r_{p/s,0}|^2$ and $|r_{p/s,-1}|^2$ from Eq. (S9) & (S10). For p-polarization, if d = 1, it indicates $\tilde{\omega}_+$ is $\tilde{\omega}_2$ ($B_1$) but is $\tilde{\omega}_4$ ($A_1$) if d = -1. On the other hand, for s-polarization, $\tilde{\omega}_+$ is $\tilde{\omega}_3$ ($B_2$) for d = 1 but $\tilde{\omega}_1$ ($A_2$) for d = -1.

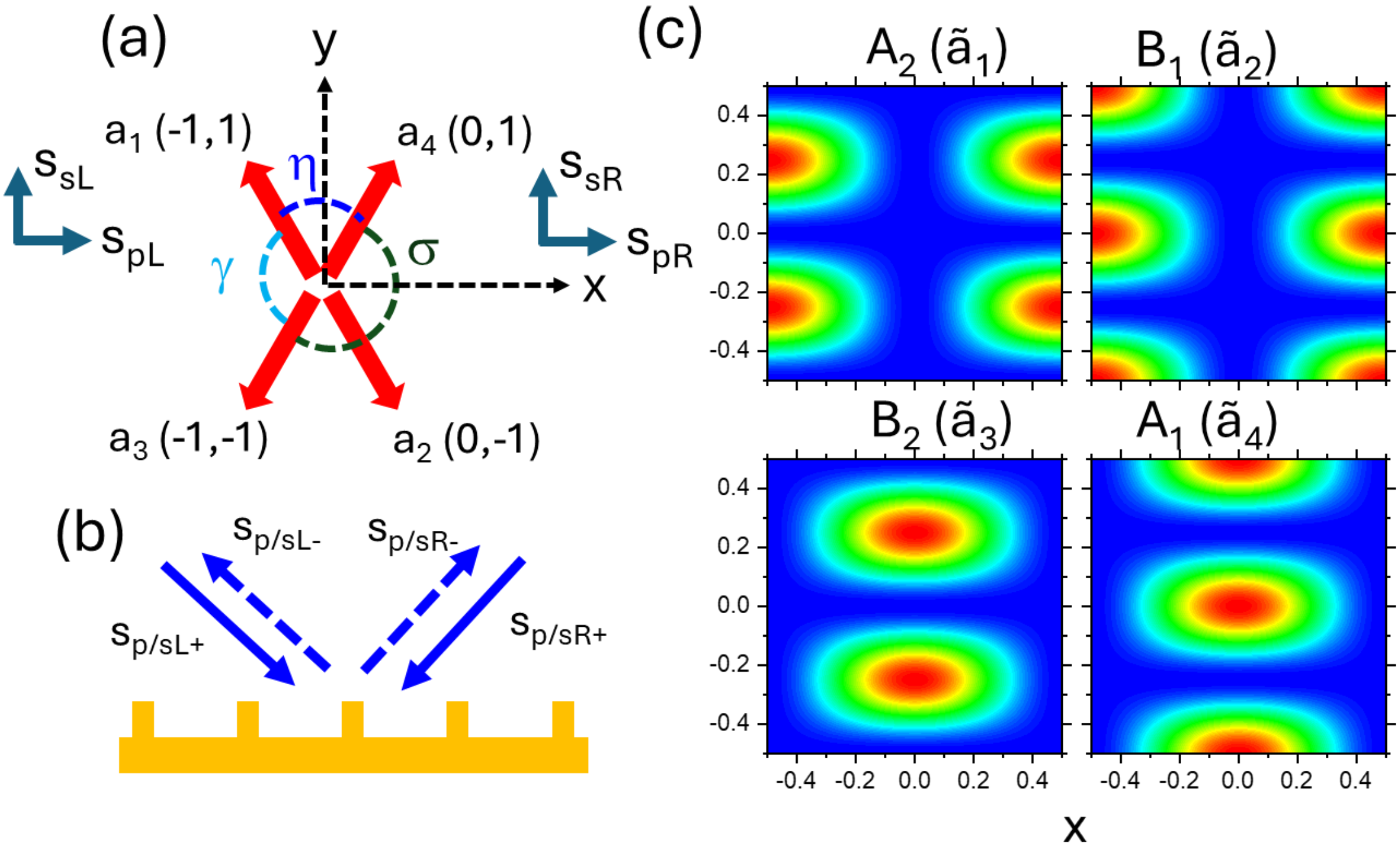

Fig. S2. (a) The schematic plane-view of how four (0,±1) and (-1,±1) SPPs couple at the X point for CMT formulation and the coupling constants are defined as η, γ, and σ. (b) The cross-section view shows the left and right incoming $s_+$ and the outgoing $s_-$ channels at the X point. Each channel carries p- and s-polarizations. If the right incoming channel is the incidence, then the left and right outgoing channels are the specular reflection and -1 diffraction order, respectively. (c) The field patterns of $\tilde{a}_{1-4}$ and are labelled as $A_2$, $B_1$, $B_2$ and $A_1$.

## C. FDTD simulation

For simulating angle- and polarization-resolved reflectivity spectra of 2D square lattice PmCs, a semi-infinite gold unit cell is used. It has same period in x- and y-directions and contains a center circular hole with radius R and hole depth H. Bloch boundary condition is used on four sides whereas perfectly matched layer is used on the top and at the bottom. The unit cell is excited by broadband plane wave under different incident polar $\theta$ and sample rotation $\varphi$ angles as well as polarization to compute the reflectivity mapping. A power monitor is placed above the unit cell to record the fields. The associated near-field patterns for any $\lambda$ can be calculated as well.

For simulating the edge states, supercell is used and it contains 20 trivial and nontrivial unit cells on the right and left sides, as shown in Fig. 10(a) in the main text. The interface is sandwiched between the trivial and nontrivial arrays. While Bloch boundary is applied in the direction perpendicular to the interface, perfectly matched layer is used in other directions. To simulate reflectivity spectra, the supercell is excited under polarized plane wave so that both $\theta$ and $\varphi$ can be controlled. On the other hand, to effectively excite the edge state, dipole excitation is used. For example, for the $k_y$-dependent radiation mapping in Fig. 11 in the main text, the linear dipoles are placed very close to the interface for predominately exciting the edge state. The outgoing fields and near-fields are then collected by power monitors.

## D. FDTD results at the Γ point

Fig. S3(a)-(d) show the p- and s-polarized total reflectivity spectral mappings taken along the Γ-X direction, i.e. $\varphi = 0^o$, at $\theta = 0^o$ and $1^o$ for different R. Fig. S4-7(a)-(c) show the reflectivity spectra taken at $\theta = 0^o$ and $1^o$ and their corresponding field patterns for R = 150, 200, 250, and 300 nm.

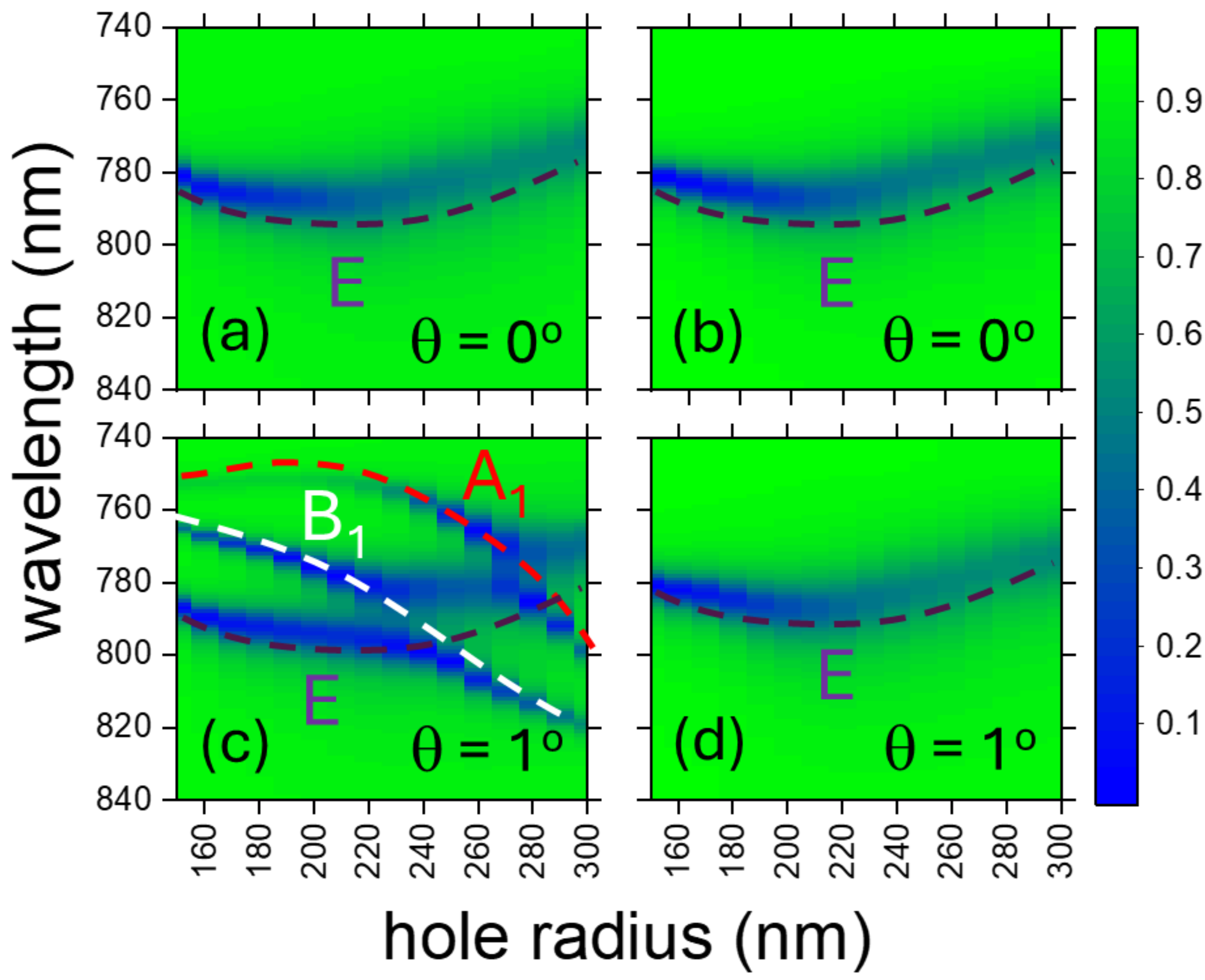

Fig. S3. The p-polarized total reflectivity spectral mappings taken along the Γ-X direction, i.e. $\varphi = 0^{o}$, at $\theta$ = (a) $0^{o}$ and (c) $1^{o}$ for different R. The s-polarized total reflectivity spectral mappings taken along the Γ-X direction, i.e. $\varphi = 0^{o}$, at $\theta$ = (b) $0^{o}$ and (d) $1^{o}$ for different R. The dash lines are the $A_1$, $B_1$, and E modes for visualization.

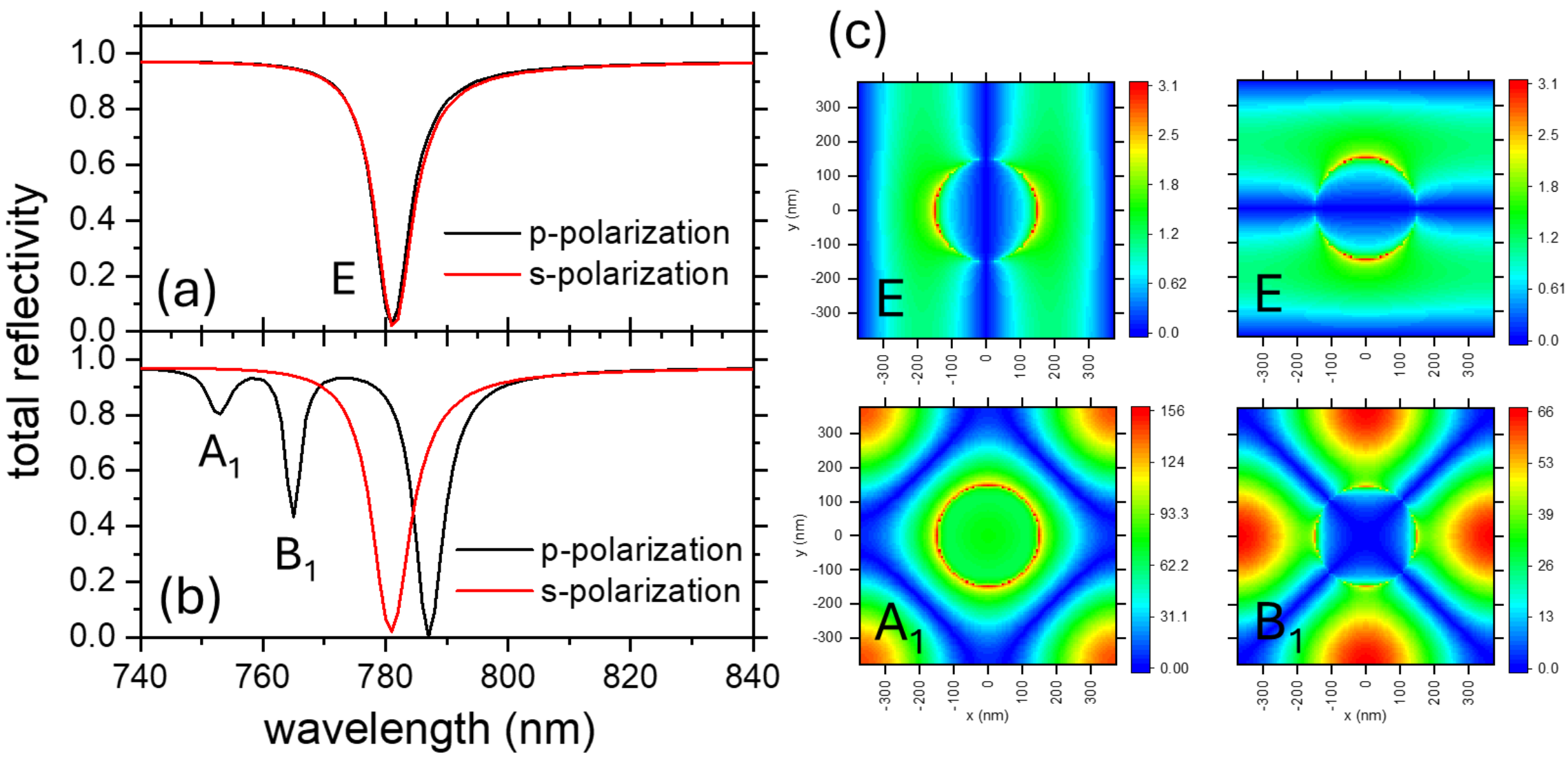

Fig. S4. For R = 150 nm, the (a) p- and s-polarized reflectivity spectra taken at the Γ point. The (b) p- and s-polarized reflectivity spectra taken at θ = 1°. (c) The corresponding electric near-field pattens taken at λ = (a) 781 nm are labelled as E and (b) = 752 and 765 nm are labelled as $A_1$ and $B_1$.

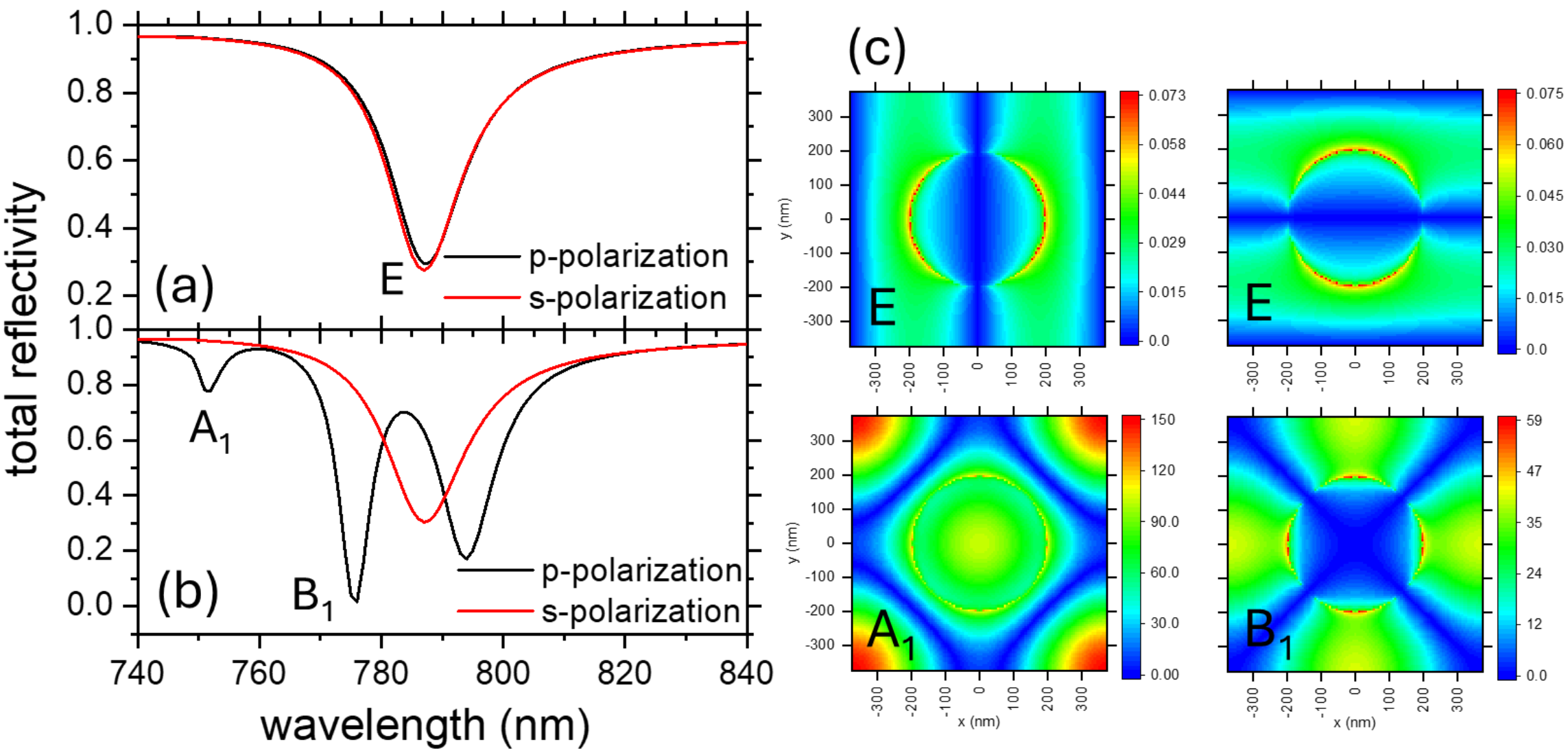

Fig. S5. For R = 200 nm, the (a) p- and s-polarized reflectivity spectra taken at the Γ point. The (b) p- and s-polarized reflectivity spectra taken at θ = 1°. (c) The corresponding electric near-field pattens taken at λ = (a) 787 nm are labelled as E and (b) = 751 and 776 nm are labelled as $A_1$ and $B_1$.

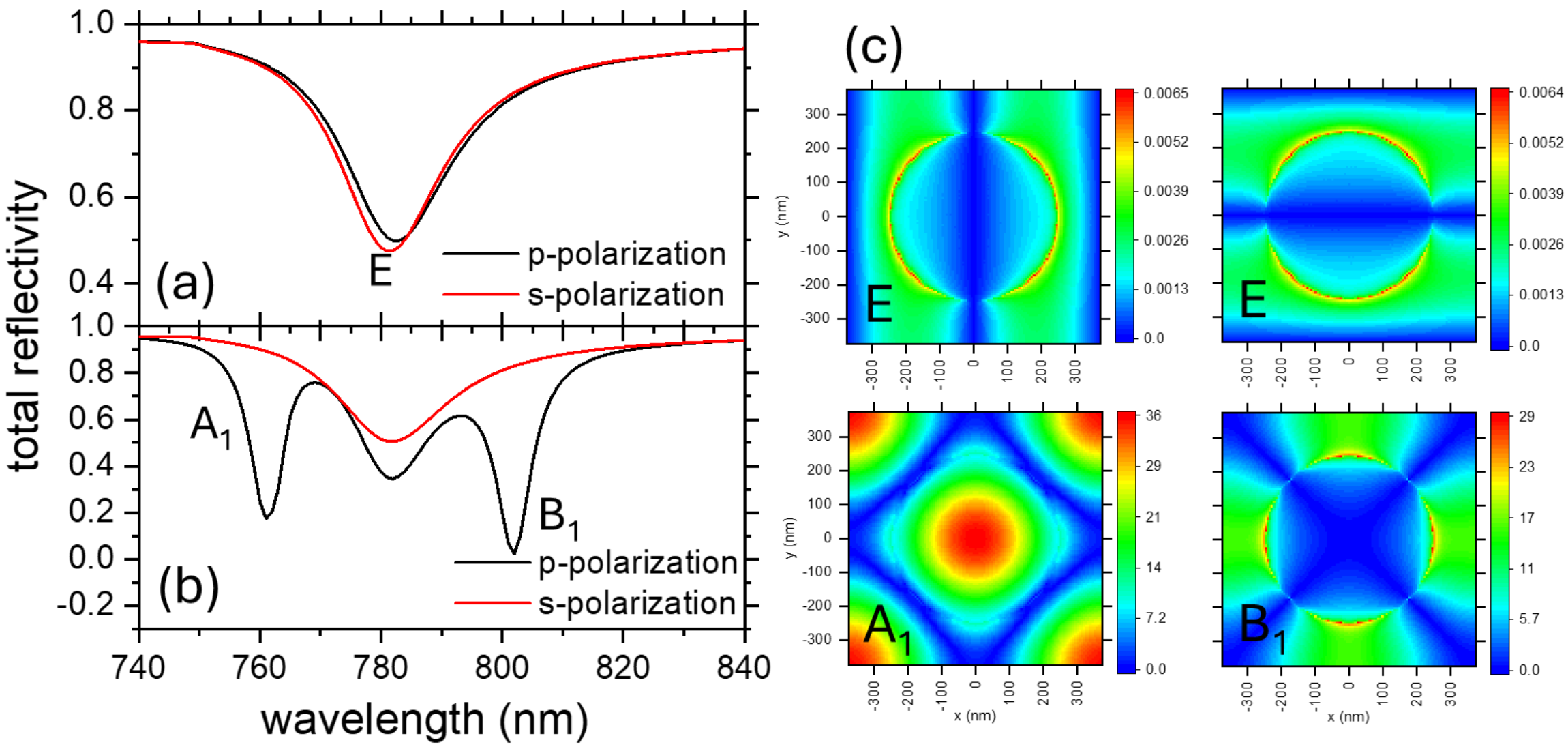

Fig. S6. For R = 250 nm, the (a) p- and s-polarized reflectivity spectra taken at the Γ point. The (b) p- and s-polarized reflectivity spectra taken at θ = 1°. (c) The corresponding electric near-field pattens taken at λ = (a) 782 nm are labelled as E and (b) = 761 and 802 nm are labelled as $A_1$ and $B_1$.

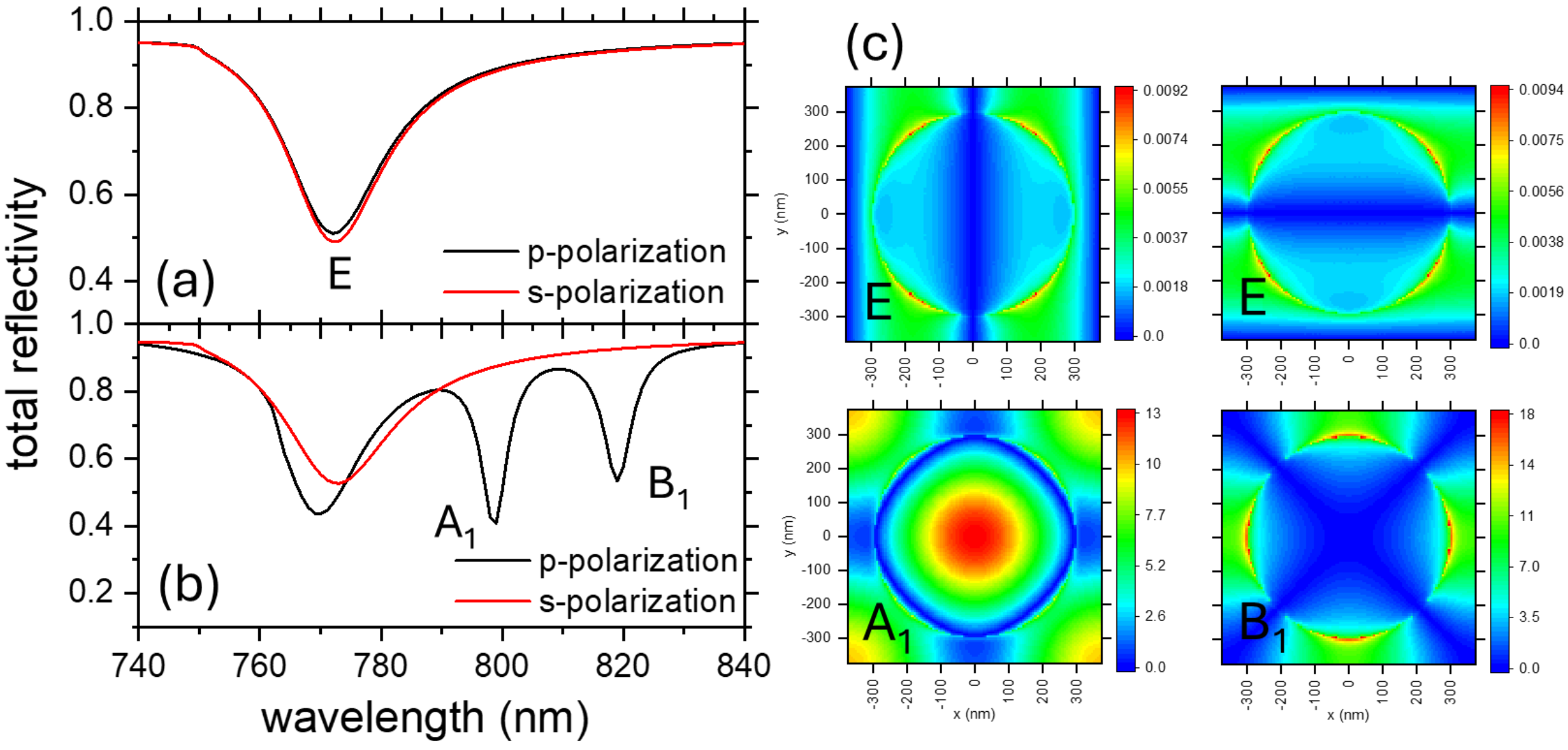

Fig. S7. For R = 300 nm, the (a) p- and s-polarized reflectivity spectra taken at the Γ point. The (b) p- and s-polarized reflectivity spectra taken at θ = 1°. (c) The corresponding electric near-field pattens taken at λ = (a) 772 nm are labelled as E and (b) = 798 and 819 nm are labelled as $A_1$ and $B_1$.

## E. FDTD results at the X point

Fig. S8(a) & (b) show the p- and s-polarized total reflectivity spectral mappings the X point for different R. Fig. S9-12(a) & (b) show the reflectivity spectra and the corresponding field patterns taken at the X point for R = 150, 200, 250, and 300 nm.

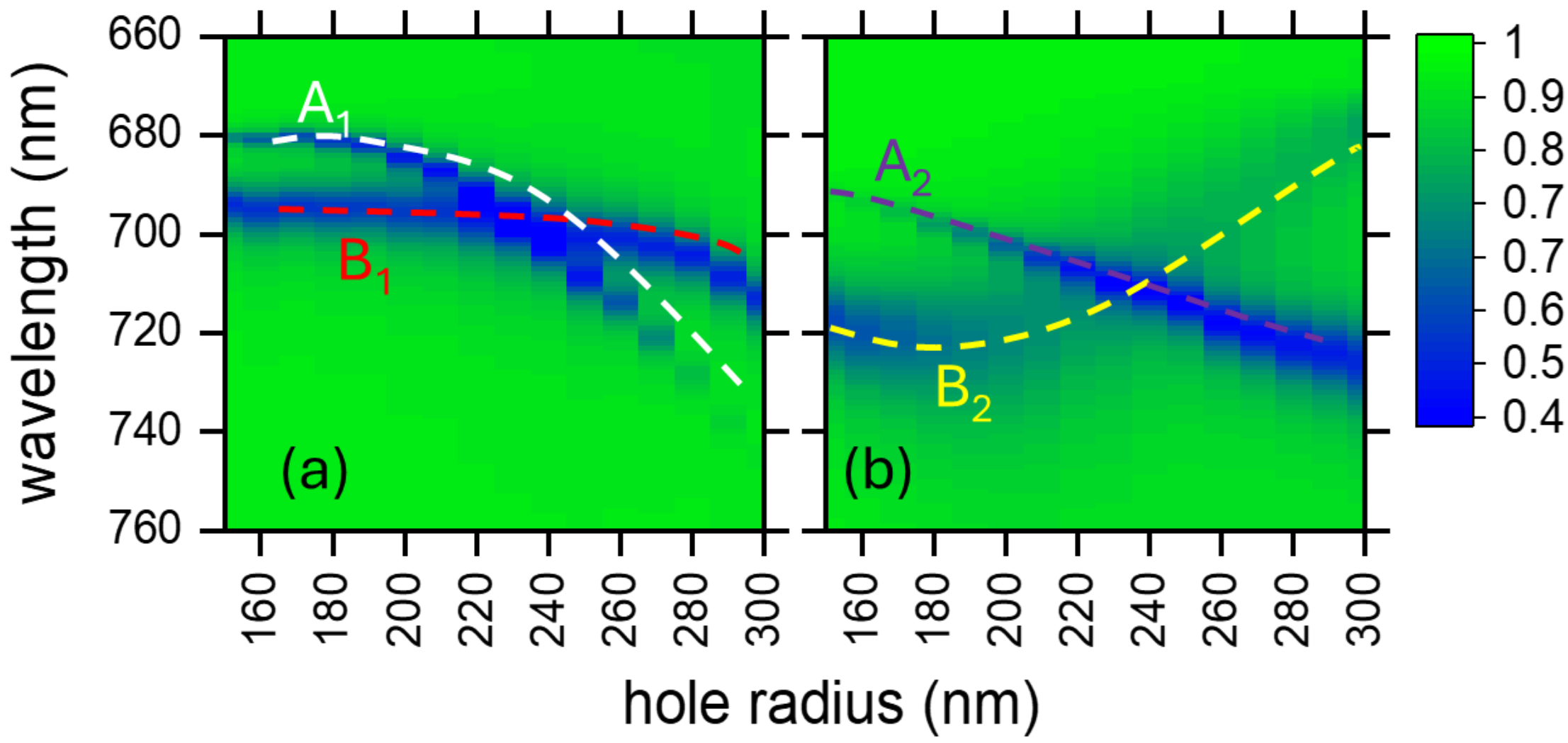

Fig. S8. The (a) p- and (b) s-polarized total reflectivity spectral mappings at the X point for different R. The color dash lines are for the visualization of the $A_1$, $B_1$, $A_2$ and $B_2$ assignments.

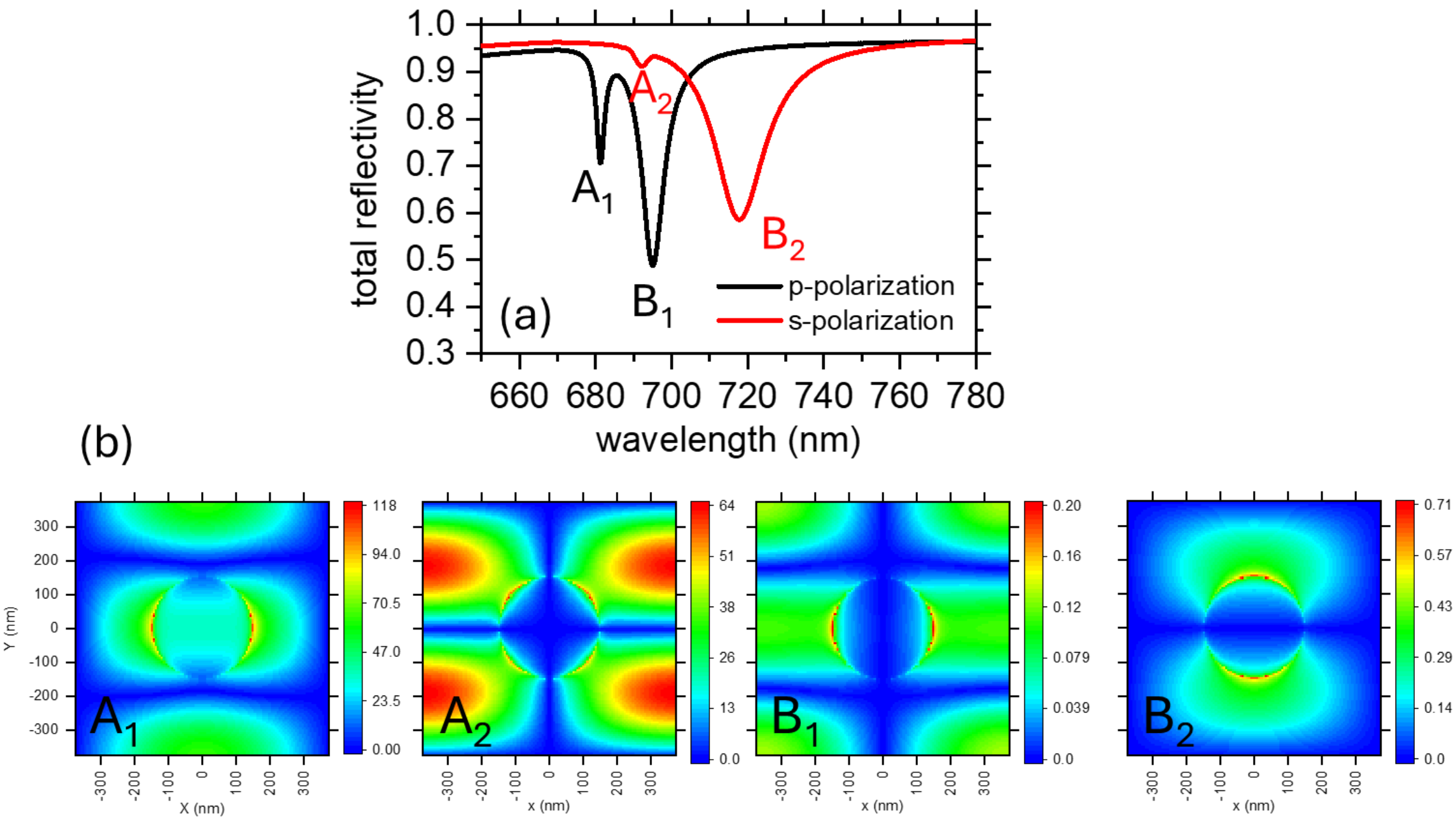

Fig. S9. For R = 150 nm, the (a) p- and s-polarized reflectivity spectra taken at the X point. (b) The corresponding field patterns at $\lambda$ = 680 (p-excited), 692 (s-excited), 694 (p-excited), and 718 (s-excited) nm and are labelled as $A_1$, $A_2$, $B_1$, and $B_2$.

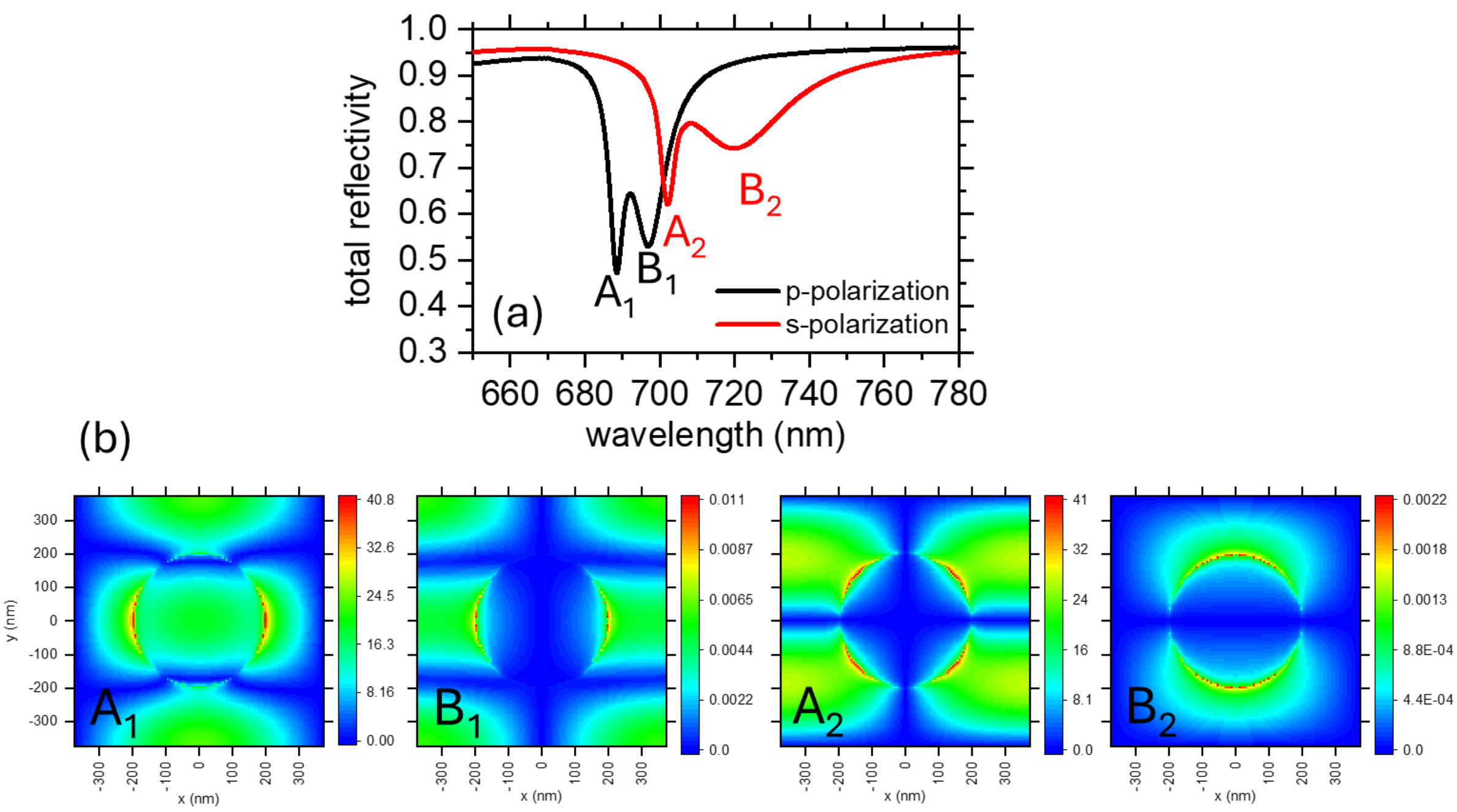

Fig. S10. For R = 200 nm, the (a) p- and s-polarized reflectivity spectra taken at the X point. (b) The corresponding field patterns at $\lambda$ = 688 (p-excited), 697 (p-excited), 702 (s-excited), and 720 (s-excited) nm and are labelled as $A_1$, $B_1$, $A_2$, and $B_2$.

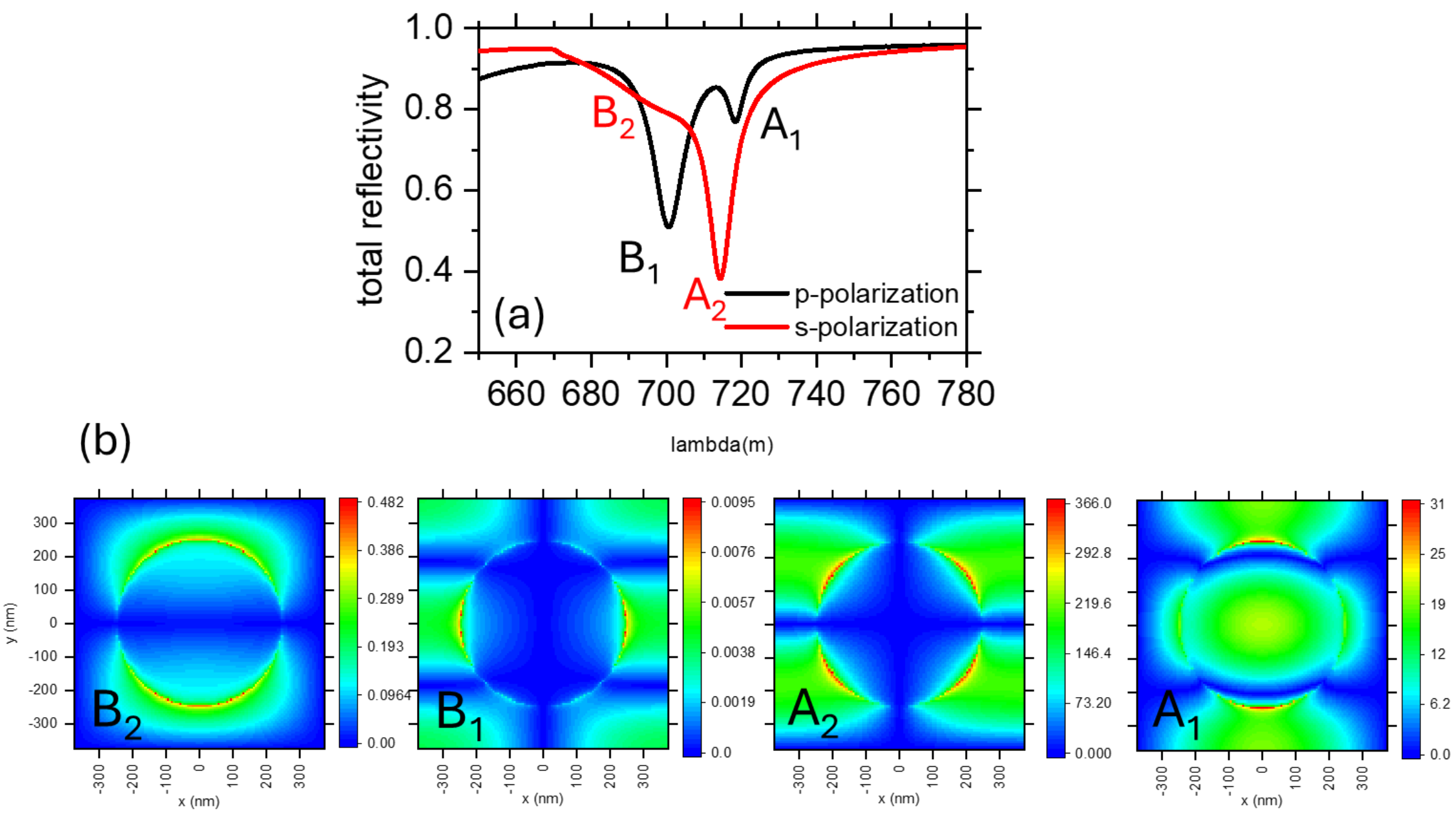

Fig. S11. For R = 250 nm, the (a) p- and s-polarized reflectivity spectra taken at the X point. (b) The corresponding field patterns at $\lambda$ = 693 (s-excited), 701 (p-excited), 714 (s-excited), and 718 (p-excited) and are labelled as $B_2$, $B_1$, $A_2$, and $A_1$.

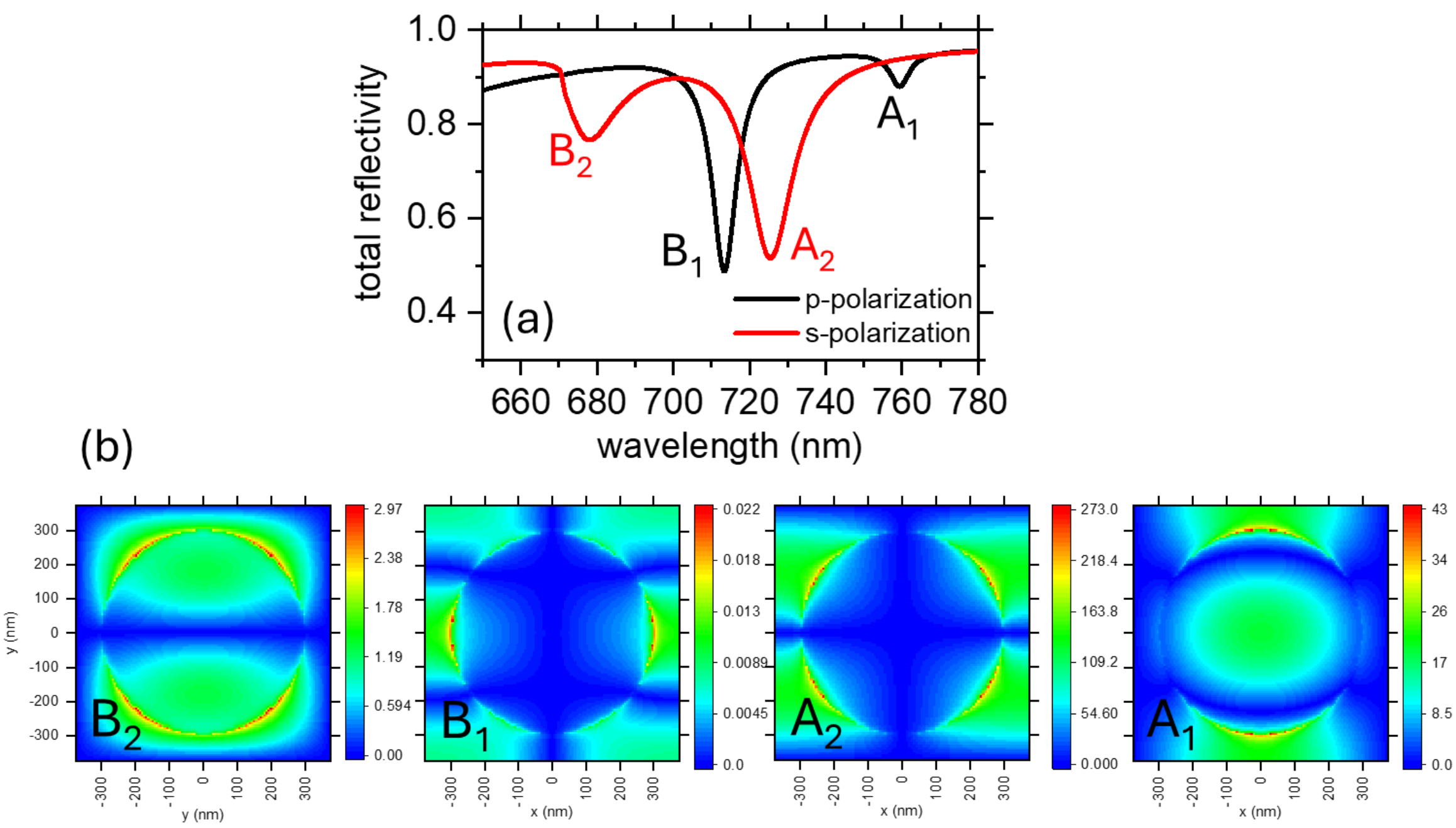

Fig. S12. For R = 300 nm, the (a) p- and s-polarized reflectivity spectra taken at the X point. (b) The corresponding field patterns at $\lambda$ = 678 (s-excited), 713 (p-excited), 725 (s-excited), and 759 (p-excited) and are labelled as $B_2$, $B_1$, $A_2$, and $A_1$.

## F. FDTD results at the M point

The dispersion relation around the M point is shown in Fig. S13. The grey area is the light cone. As the modes at the M point are below the light cone, they are not accessible by far-field. Therefore, dipole excitation is used for calculating the p- and s-polarized radiation spectral mappings as a function of R.

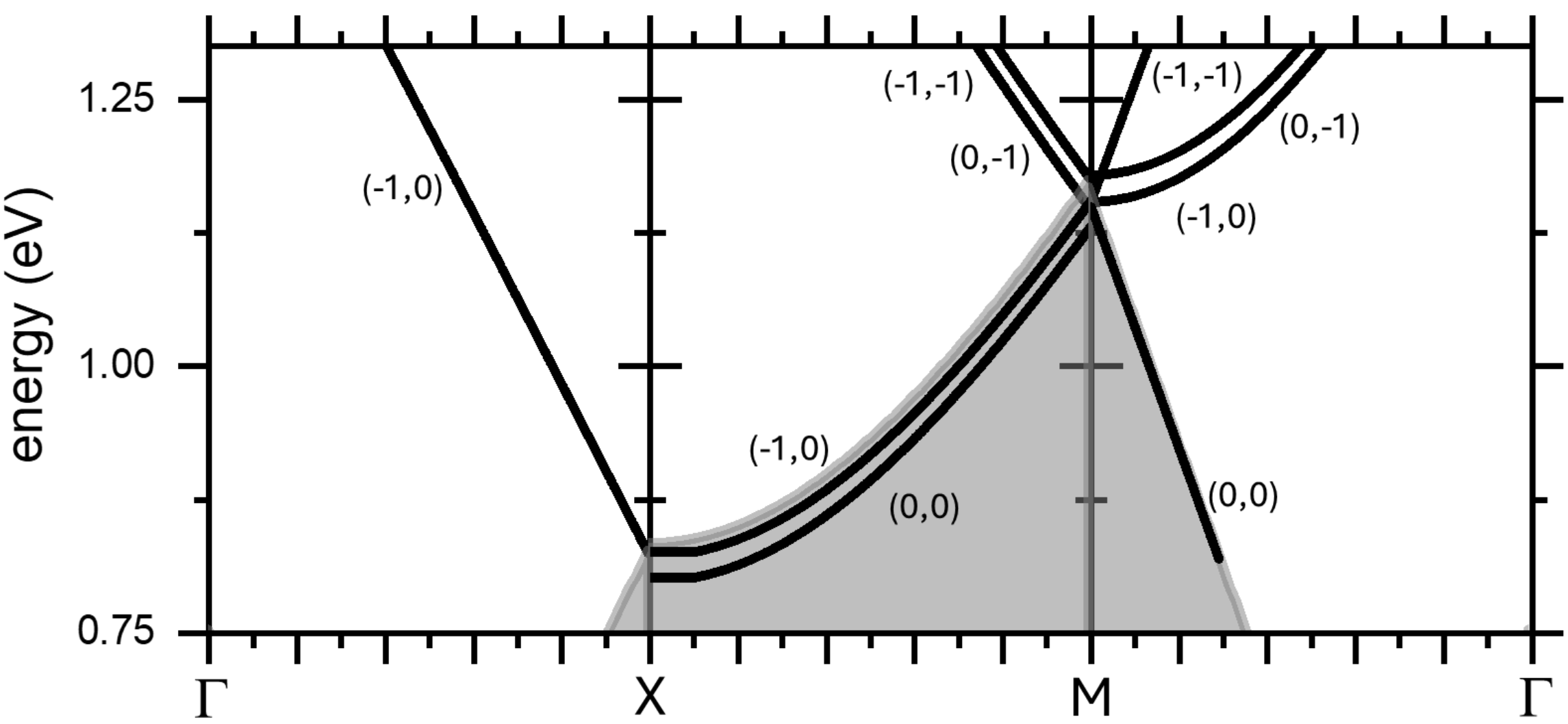

Fig. S13. The dispersion relation of SPP around the M point sue to the coupling of four (0,0), (-1,0), (0,-1) and (-1,-1) modes

Fig. S14(a) & (b) show the p- and s-polarized radiation spectral mappings taken at the M point for different R. All four nonradiative modes are seen, and they resemble to the results of the Γ point. Two degenerate modes and are assigned as E whereas other two nondegenerate are identified as $A_1$ and $B_2$ following the field patterns illustrated in Fig. S15-17 for R = 150, 200, and 250 nm. We notice that E and $B_2$ are about to merge at R = 300 nm, possibly yielding a Dirac point and band inversion at larger R.

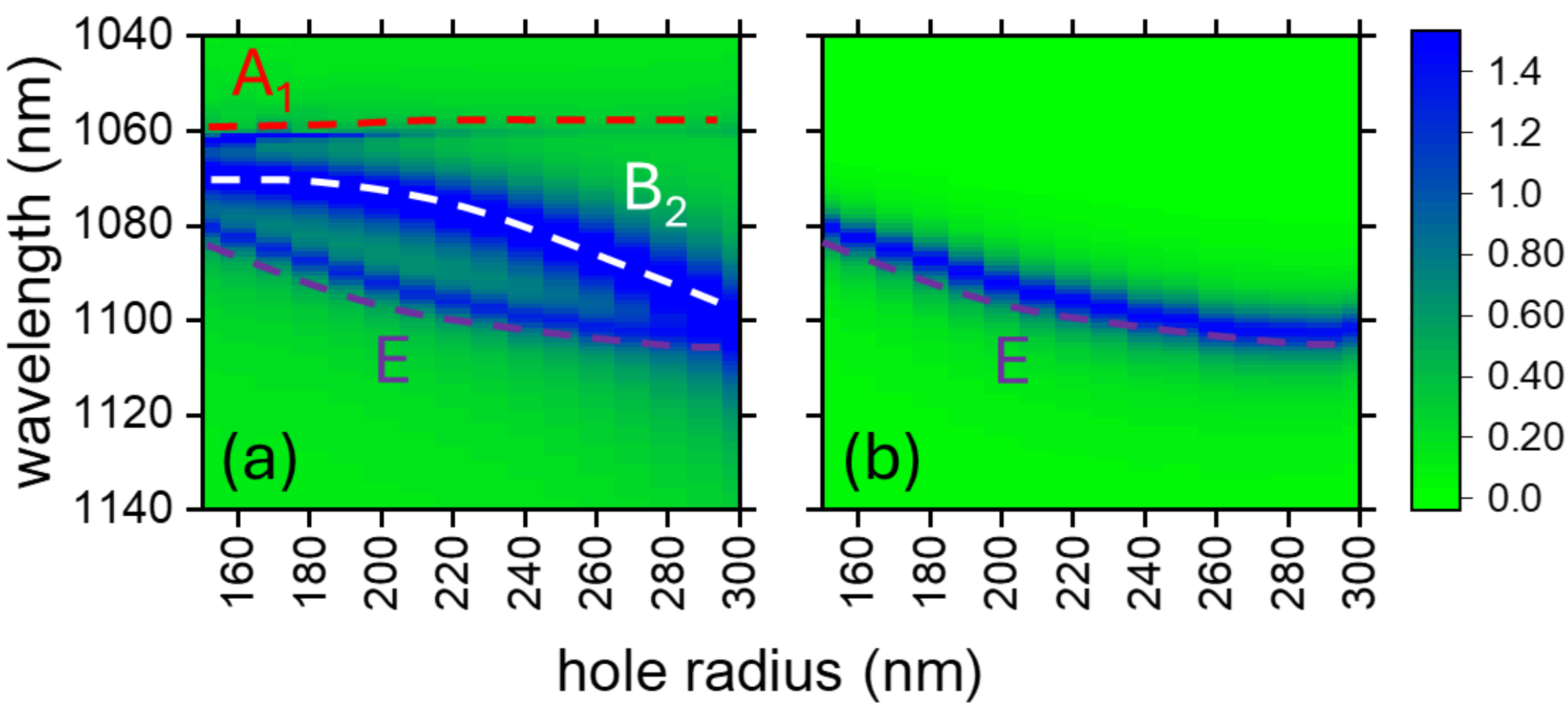

Fig. S14. The (a) p- and (b) s-polarized radiation spectral mappings for different R. The color dash lines are for the visualization of the $A_1$, $B_2$, and E assignments.

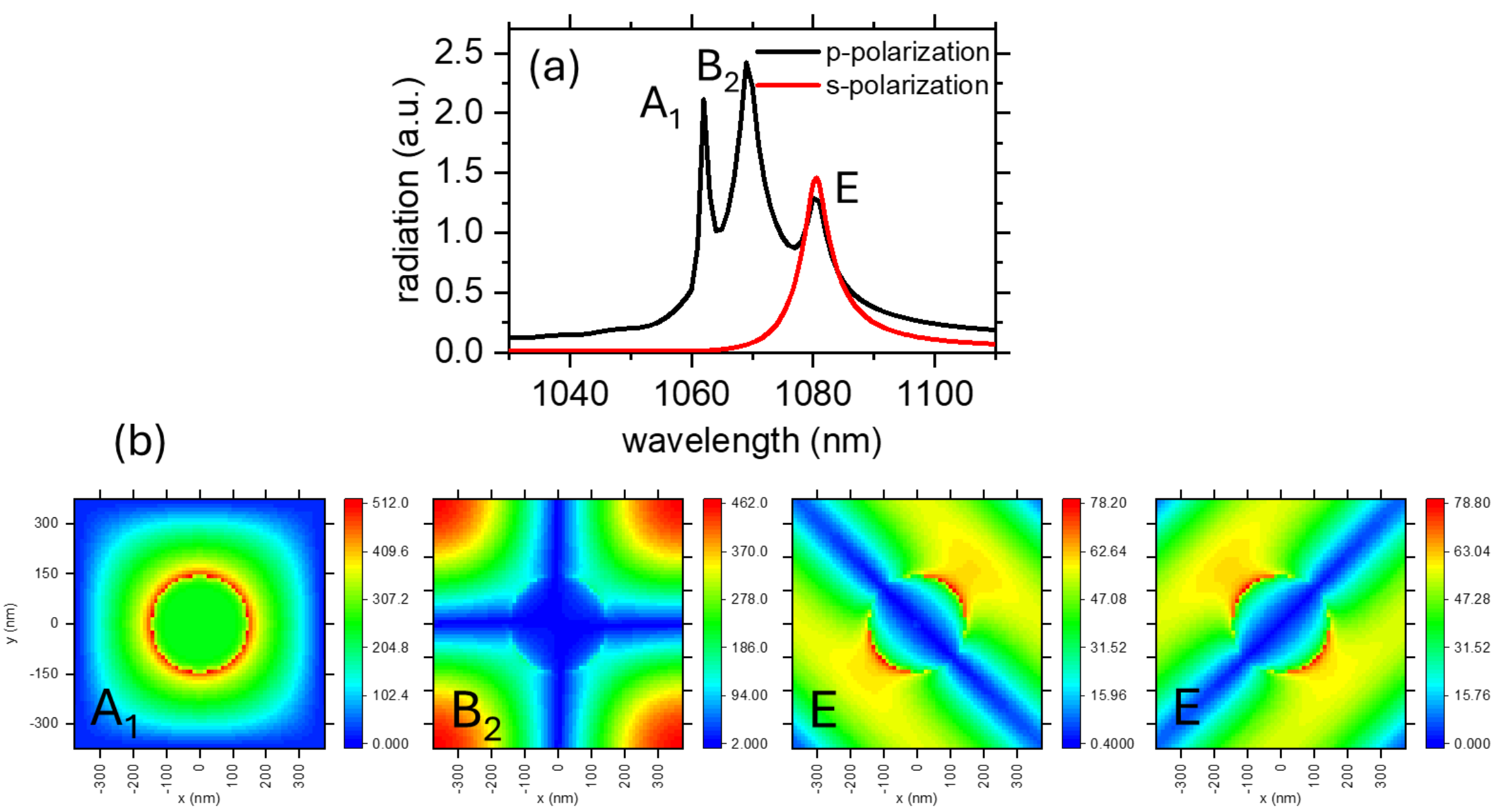

Fig. S15. For R = 150 nm, the (a) p- and s-polarized radiation spectra. (b) The corresponding field patterns at λ = 1062, 1069, and 1080 nm and are labelled as $A_1$, $B_2$, and E.

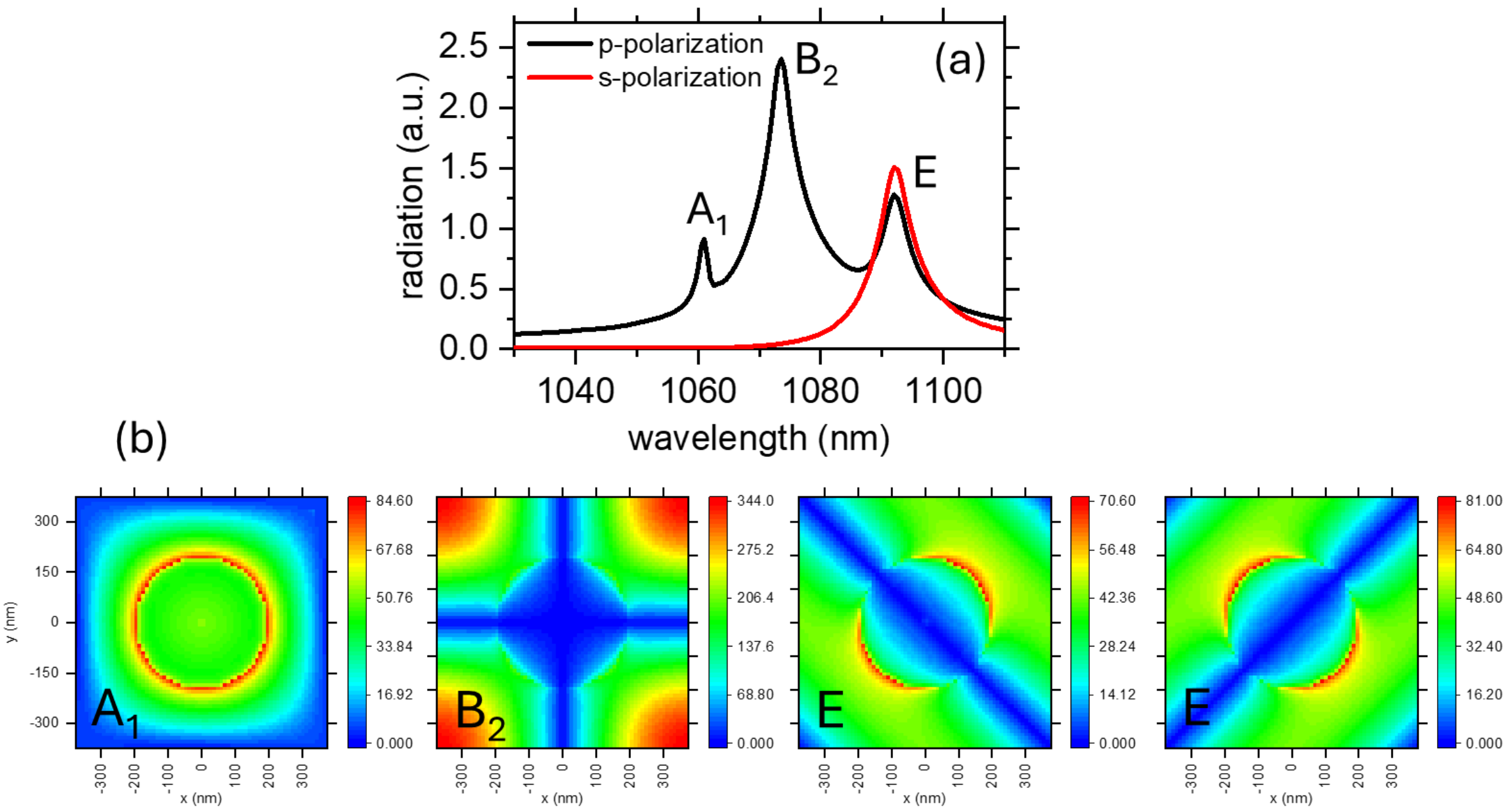

Fig. S16. For R = 200 nm, the (a) p- and s-polarized radiation spectra. (b) The corresponding field patterns at λ = 1060, 1073, and 1092 nm and are labelled as $A_1$, $B_2$, and E.

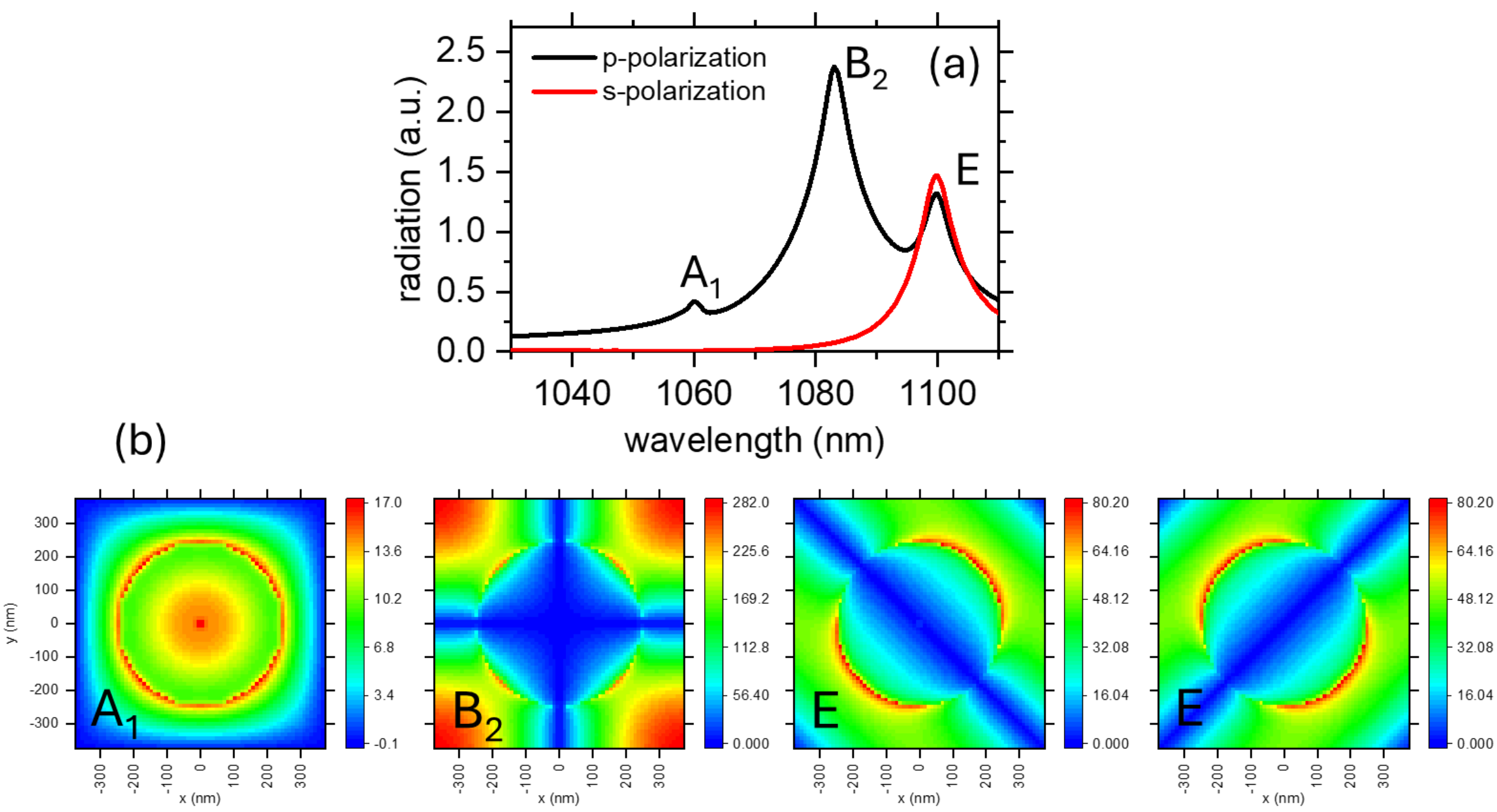

Fig. S17. For R = 250 nm, the (a) p- and s-polarized radiation spectra. (b) The corresponding field patterns at λ = 1060, 1083, and 1100 nm and are labelled as $A_1$, $B_2$, and E.

### G. The field patterns at different ($\lambda$,$k_y$) for the lower and upper bands

Fig. S18 is the $k_y$-dependent radiation spectral mapping excited by linear dipole, which is placed at the interface. Different ($\lambda$,$k_y$) positions of the upper and lower bands are labeled and their corresponding field patterns in logarithmic scale are illustrated in Fig. S19(g) – (m). We see the lower band, Fig. S19(h), (j) & (k), are the bulk states where the field strengths spread over the bulk region. On the other hand, for the upper band in Fig. S19(g), (i), (l) & (m), begins as an edge state but is mixed with bulk states when $k_y$ increases

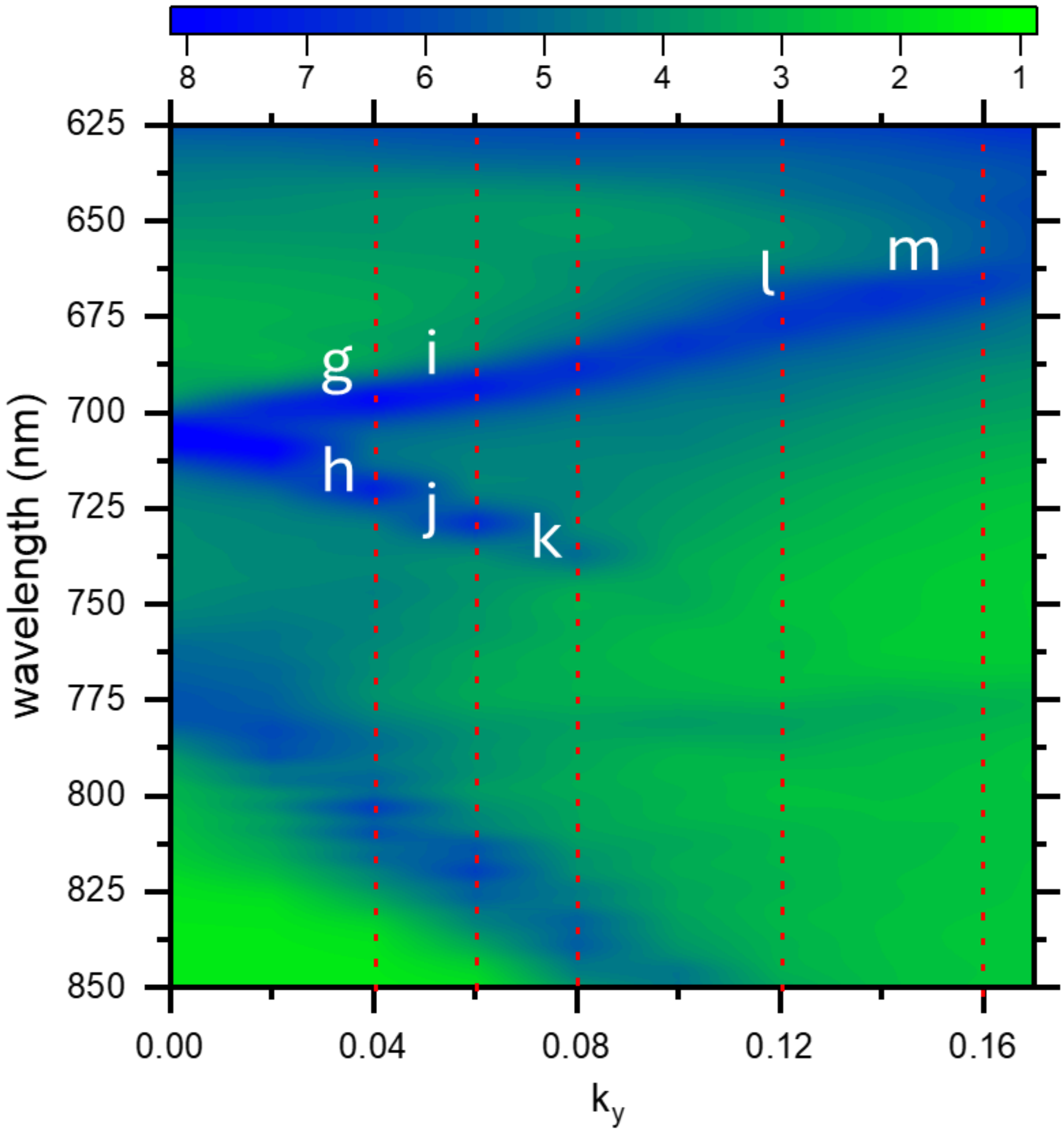

Fig. S18. The $k_y$-dependent radiation spectral mapping taken under dipole excitation.

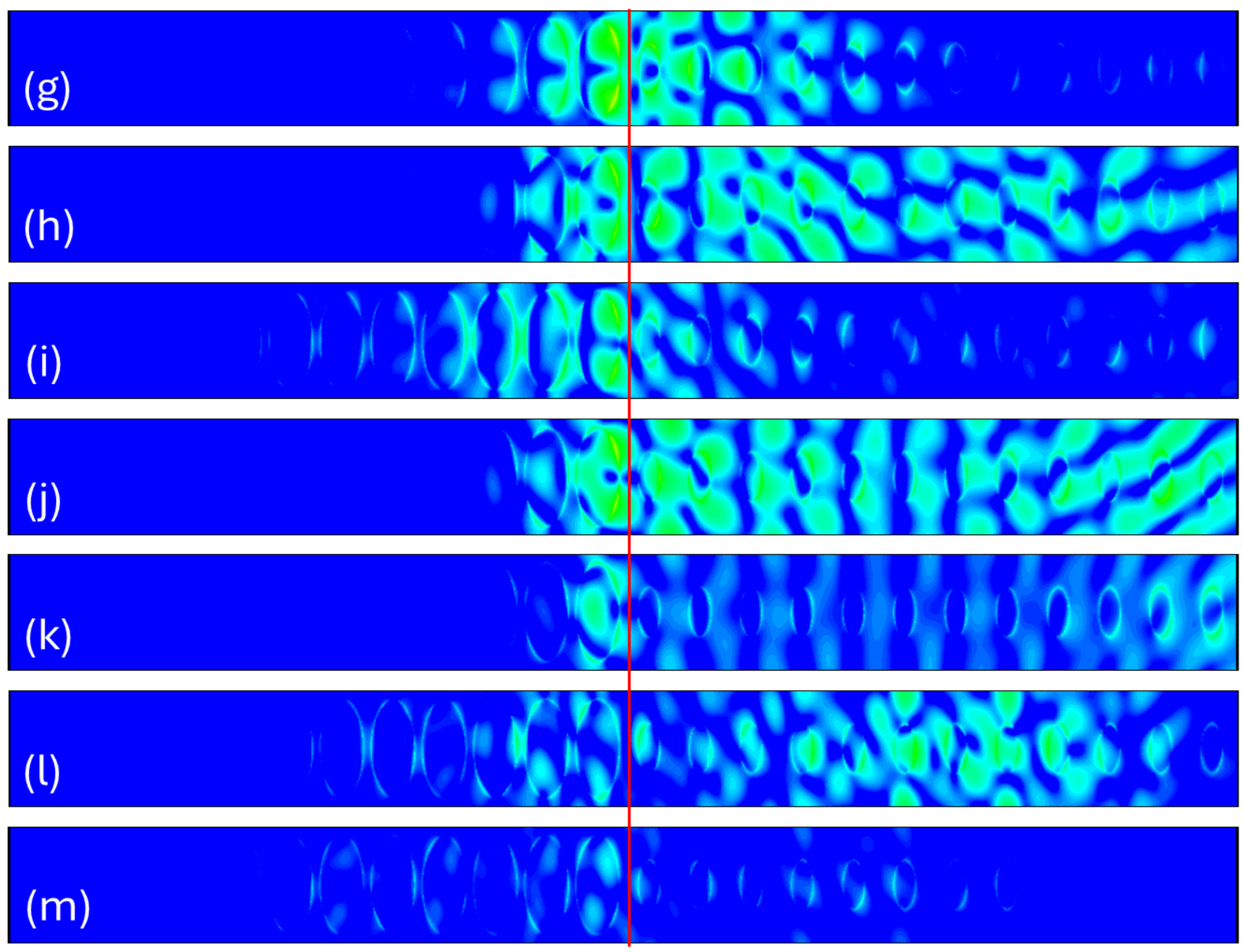

Fig. S19. The field patterns in logarithmic scale at $(\lambda,k_y)$ = (g) (697,0.04), (h) (720,0.04), (i) (693,0.06), (j) (729.0.06), (k) (737,0.08), (l) (677,0.12) and (m) (666,0.16). The dash line indicates the interface.

## H. Schematic of the Fourier space optical microscope for angle- and wavelength resolved diffraction mapping

Fig. S20 shows the schematic of the Fourier space optical microscope. Briefly, a broadband supercontinuum laser from a nonlinear photonic crystal fiber is collimated and then passed through a set of linear polarizers, wave plates, and lenses before being focused onto the back focal plane (BFP) of a 100X objective lens (OB) with numerical aperture = 0.9. The light exiting from the objective lens is then a collimated beam with well-defined linear polarization. In addition, by displacing the focused spot across the BFP of the objective lens using a motorized translation stage, the incident polar angle $\theta_i$ of the collimated beam onto the sample can be varied following $\sin\theta_i = d/f$, where $d$ is the distance between the focused spot and the optical axis of the BFP and $f$ is the focal length of the objective lens. In addition, the azimuth angle $\phi$ can be varied by a motorized rotation sample stage to align the incident plane to the Γ-X direction of the PmC. The diffractions from the PmC are then collected by the same objective lens and are routed through a set of lens system so that the diffraction orders are projected onto the Fourier space. By placing an aperture at the Fourier space to filter out the desired diffraction order, its intensity and phase spectra can be measured by a spectrometer-based CCD detector. The PmCs prepared by FIB have around 25 periods on one side, giving an area around $19 \times 19$ $\mu m^2$. The beam size was around $10 \times 10$ $\mu m^2$, covering around 200 unit cells.

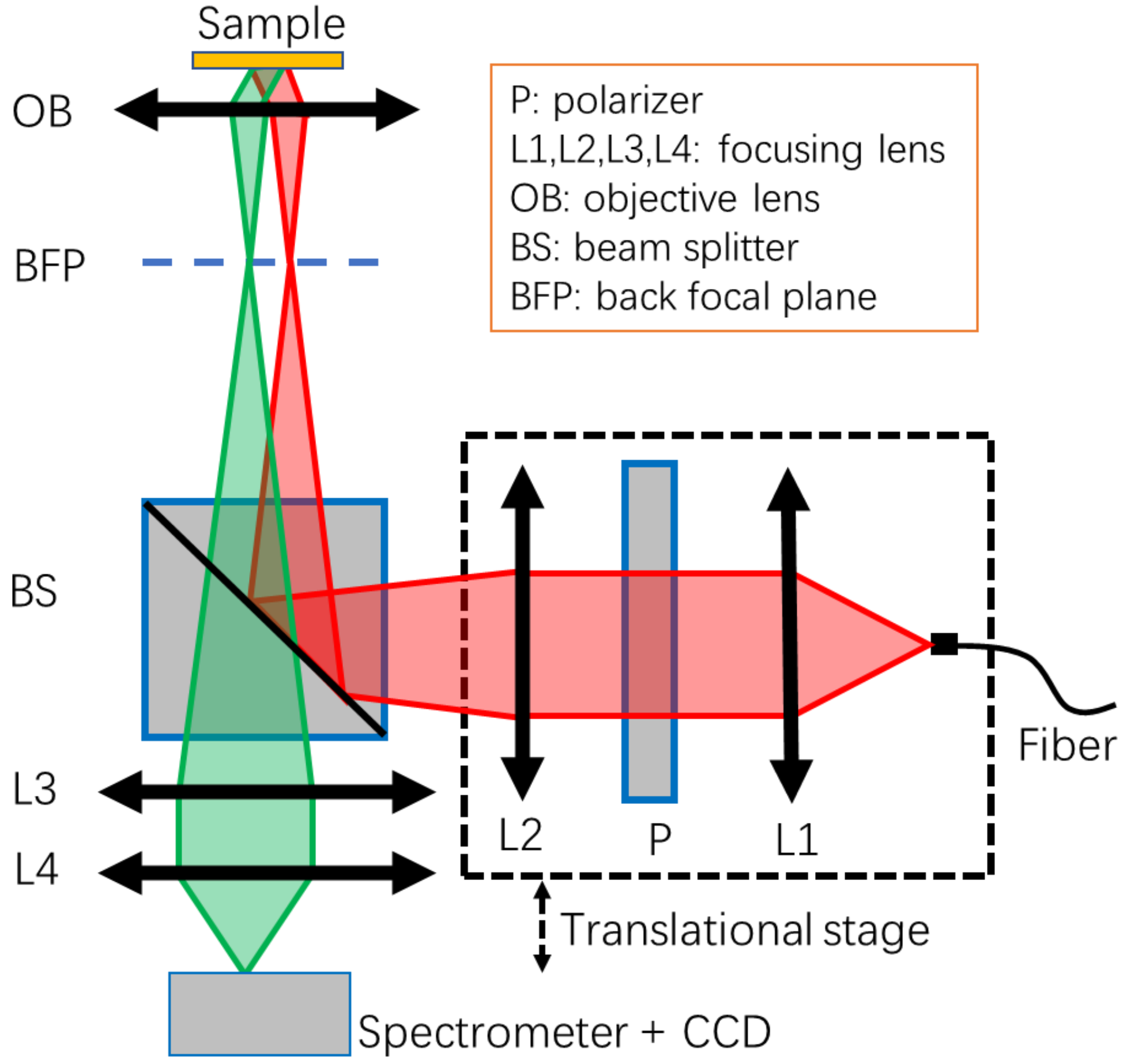

Fig. S20. The schematic of the Fourier optical microscope.